\documentclass[journal]{IEEEtran}
\usepackage[table]{xcolor}

\usepackage{todonotes} 

\usepackage{optidef}
\usepackage{subcaption}
\usepackage{graphicx}
\usepackage[export]{adjustbox}
\usepackage{tabularx,adjustbox}
\usepackage{amsmath,amssymb,bm,mathrsfs}
\usepackage{siunitx}

\usepackage{breqn}
\usepackage{graphicx}
\usepackage{mwe}
\usepackage{arydshln}
\usepackage{makecell}
\usepackage{adjustbox}
\usepackage{float}
\usepackage{nomencl}
\makenomenclature
\usepackage{mathtools}
\usepackage{bm}
\usepackage{soul}
\usepackage{siunitx}
\usepackage{cite}
\usepackage{multirow}
\usepackage{arydshln}
\usepackage{xfrac}
\usepackage{nicefrac}
\usepackage{nicematrix,tikz}
\usepackage[T1]{fontenc}
\usepackage{inputenc}
\usepackage{babel}
\usepackage[font=small,labelfont=bf]{caption}
\DeclareSIUnit \var { var } 
\usepackage{titlesec}
\usepackage{environ}         
\usepackage{etoolbox}        
\usepackage{graphicx}        
\usepackage{hyperref}
\usepackage{footnote}
\usepackage{comment}

\RenewEnviron{equation}{
  \settowidth{\myl}{$\BODY$}                       
  \origequation
  \ifdimcomp{\the\linewidth}{>}{\the\myl}
  {\ensuremath{\BODY}}                             
  {\resizebox{0.89\linewidth}{!}{\ensuremath{\BODY}}}  
  \origendequation
}

\newcommand{\overbar}[1]{\mkern 1.5mu\overline{\mkern-1.5mu#1\mkern-1.5mu}\mkern 1.5mu}

\begin{document}

\title{Grid-Aware Islanding and Resynchronisation of AC/DC Microgrids}

%
%
%

\author{Willem~Lambrichts,~\IEEEmembership{Student Member,~IEEE,}
        and Jules~Mace,~\IEEEmembership{Member,~IEEE,}
        and Drazen~Dujic,~\IEEEmembership{Fellow,~IEEE,}
        and Mario~Paolone,~\IEEEmembership{Fellow,~IEEE,}
\thanks{The authors are with the Swiss Federal Institute of Technology of Lausanne, Switzerland, email: \{willem.lambrichts,jules.mace,drazen.dujic,mario.paolone\}@epfl.ch. The project has received funding from the European Union’s Horizon 2020 Research \& Innovation Programme under grant agreement No. 957788.}}%

\maketitle

\begin{abstract}

This paper proposes an optimal, grid-aware control framework for the islanding, island-operation and resynchronisation of hybrid AC/DC microgrids. 

The optimal control framework is based on a formally derived linearized load-flow model for multiterminal hybrid AC/DC networks. 
The load flow model integrates the AC grid, DC grid, and interfacing converters (IC) into a unified representation. This work extends an existing load flow model to include the ICs' grid-forming operation.

In traditional islanding control frameworks, the grid-forming converter is typically interfaced with an energy storage system that can provide bidirectional power to maintain the power balance. The proposed framework, however, allows the ICs to operate as the grid-forming unit while being connected to a DC grid rather than a single resource. This configuration allows for a wider operating range and, thus, a more flexible control. 
Furthermore, the optimal grid-aware control framework can steer the system to ensure a feasible operation without any grid constraint violations before, during, and after the islanding manoeuvre. The framework also guarantees smooth transitions, i.e., without any significant transient behaviour, when transitioning between grid-connected and islanding operations. 
The optimal control framework is experimentally validated on a 27-bus hybrid AC/DC network consisting of 3 ICs that interface the AC and DC networks. The hybrid grid hosts various controllable and stochastic resources. 


\end{abstract}








\vspace{-15pt}
\section{Introduction}
\label{Introduction}

Islanded operation of segments of distribution systems can provide distinct benefits to the overall system performance, particularly regarding the reliability and quality of service, provided it is managed and coordinated efficiently \cite{313}. 
More than three-quarters of grid failures occur in medium and high-voltage systems \cite{421}. Therefore, driven by the increasing penetration of distributed generation at distribution levels, the opportunity increases for islanded operation of microgrids subsequent to failures in the medium and high-voltage grid. 
This can greatly improve the consumers' uninterrupted supply and reduce the lost customer minutes \cite{421}. Reference \cite{celli2005distributed} demonstrates that intentional islanding manoeuvres can improve the overall reliability when compared to conventional approaches that rely on redundant supply paths. 


Additionally, most of the distributed energy resources (DER) in the distribution system have a DC nature. Therefore, rather than employing individual DC-AC converters, these DER can be directly connected to a DC grid. This DC grid is interfaced with the AC grid at one or more strategically selected nodes to create hybrid AC/DC grids. This configuration allows for more flexible operation (facilitated by controllable AC-DC interfacing converters (ICs)) 
\cite{BookACDCcontrol}, reduces the overall cost of the system, and increases global efficiency due to the need for fewer power conversion stages 
\cite{BookACDCcontrol, eghtedarpour2014power}.

Most methods developed in the literature for the islanding and resynchronisation of hybrid AC/DC networks rely on myopic control strategies, such as droop control. 
The authors in \cite{ding2014control, antalem2022decentralized, sajid2019control, liang2019coordination} all present droop-based \textit{V/f} controls for the ICs. The droop-based control is experimentally validated in \cite{antalem2022decentralized} using a hardware-in-the-loop platform. Reference \cite{cai2021research} proposes an improved droop control that limits fluctuations in frequency and DC voltage during transitions.
References \cite{wang2021improved} and \cite{zhang2023seamless} present a control method based on droop control and introduce a pre-synchronisation phase to ensure smooth state transitions. 
Another myopic control method for islanded AC/DC grids is presented in \cite{li2020novel}. The method improves the dynamic response of the IC by supplying virtual inertia and damping. An energy storage system (ESS) is required in both the AC and DC grids. 
Reference \cite{melath2019novel} presents a new control method for hybrid AC/DC networks that incorporates a synchronous generator. It aims to improve transient behaviour by utilising the inertia of the generator. The authors provide an experimental validation on a small hybrid system. 
In reference \cite{qu2023control}, a state-tracking control between P and V controls is presented that ensures a smooth transition from grid-connected to islanding mode with a minimal DC voltage drop. 

These droop-based control strategies are typically easier to implement and scale. However, as is known, these controls are myopic regarding the grid's operational limits. Therefore, these methods cannot guarantee that the grid constraints will not be violated during operation. 
Another major difference between the methods presented in the literature is the network configuration.
Typically, the grid-forming converter is connected to an ESS, which becomes the AC slack bus during islanded operation. This bus regulates the voltage magnitude and frequency of the AC grid, and its active and reactive power injections result from the power balance across the entire islanded grid. Meanwhile, in the DC grid, a DER interfaced by a voltage-controllable converter regulates the DC voltage.

However, in the proposed framework, the grid-forming converter can be one of the converters that interface the AC and DC grids. This fundamentally changes the power flow problem because, during island operation, this IC injects power to maintain the active power balance in the AC grid, and the resulting active power is supplied by the DC grid. 
Consequently, the slack node for the active power is physically situated in the DC grid corresponding to the voltage-controllable converter, whereas the slack node for reactive power remains on the AC side of the grid-forming IC. This configuration is fundamentally different from previously proposed methods. It provides greater flexibility but requires the development of new computational methods to model this grid-forming IC.

In this respect, this work presents a grid-aware optimal control framework for the islanding, islanded operation and resynchronization manoeuvres of hybrid AC/DC networks.
The framework is based on the sensitivity coefficient (SC)-based optimal power flow (OPF) formulation to compute the optimal setpoints for the controllable DERs and ICs, ensuring a secure and optimal operation without any grid violations throughout the full islanding and resynchronisation manoeuvre. Furthermore, the framework's state machine ensures that the boundary conditions of the system remain the same before and after the state transitions. This guarantees a seamless islanding and resynchronisation manoeuvre without significant transient behaviour. 

The SCs are partial derivatives of the nodal voltages and branch currents with respect to the decision variables, typically controllable nodal power injections. They are calculated analytically \cite{christakou2015real} based on a unified load flow model for hybrid AC/DC networks, originally presented in 
\cite{willem_LF}. This load flow model, however, only includes ICs operating in grid-following mode. Therefore, this work presents an extension to incorporate the grid-forming operation of these ICs. Additionally, a detailed loss model of the ICs is presented to improve the accuracy of the SC-based OPF. 

The SC-based OPF results in a quadratic program (QP) that can be solved efficiently, allowing for fast (i.e. sub-second) control actions. This is particularly interesting in time-critical control applications such as islanding manoeuvres when addressing the intermittency of renewables.

\vspace{-5pt}
\section{Unified AC/DC grid model}
\label{Sec:Model}

The optimal control framework for the island and re-synchronisation manoeuvres is based on a unified load flow model that integrates the AC grid, DC grid, and ICs. The load flow model allows for various IC operating modes. The model is based on reference \cite{willem_LF} and has been suitably extended here to incorporate an accurate IC loss model and the grid-forming operation of the ICs\footnote{The source code is made available on \url{https://github.com/DESL-EPFL}}. 
\vspace{-10pt}
\subsection{Unified load flow model}
\begin{figure}[h]
 \centering
   \includegraphics[width=\linewidth]{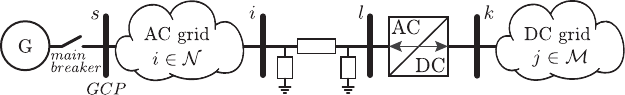}
   \caption{The generic hybrid AC/DC network. For simplicity, only one AC-DC converter is illustrated. The Grid Connection Point (GCP) and the main breaker are indicated.}
   \label{fig:gengrid} \vspace{-10pt}
 \end{figure}

Consider a generic hybrid AC/DC network illustrated in Figure \ref{fig:gengrid}. The AC grid consists of $ i \in \mathcal{N}$ nodes, the DC grid of $j \in \mathcal{M}$ nodes and both grids are interfaced by one or more ICs. The ICs are connected in buses $(l, k) \in \mathcal{L}$ to the AC and DC grids, where $l \in \mathcal{N}$ and $k \in \mathcal{M}$. 
\noindent The generic three-phase hybrid AC/DC grid is described in \eqref{eq:YY}, 
\small
\begin{align}
    \begin{bmatrix} 
        \overbar{\mathbf{I}}^{ac} \\ 
        \mathbf{I}^{dc} 
    \end{bmatrix} = 
    \begin{bmatrix} 
        \overbar{\mathbf{Y}}^{ac} & 0 \\ 
        0 & \mathbf{Y}^{dc} 
    \end{bmatrix} 
    \begin{bmatrix} 
        \overbar{\mathbf{E}}^{ac} \\ 
        \mathbf{E}^{dc} 
    \end{bmatrix}, \label{eq:YY}
\end{align}
\normalsize
where $\overbar{\mathbf{E}}^{ac}$ and $\mathbf{E}^{dc}$ are the AC and DC phase-to-ground nodal voltage vectors, $\overbar{\mathbf{I}}^{ac}$ and $\mathbf{I}^{dc}$ the AC and DC nodal current injection vectors and $\overbar{\mathbf{Y}}^{ac} = \mathbf{G}^{ac} + j \mathbf{B}^{ac}, \  \in \mathbb{C}^{\lvert \mathcal{N} \rvert }$ and $\mathbf{Y}^{dc} = \mathbf{G}^{dc}, \ \in  \mathbb{R}^{\lvert \mathcal{M} \rvert}$ the compound admittance matrices of the AC and DC grid, respectively, which are assumed to be known.

The AC grid is modelled using the standard load flow theory. It is characterised by slack nodes, PQ nodes, and PV nodes depending on the controllable variables (see Table \ref{NodeTypes}). 

The DC grid is modelled identically to the AC grid, where the electrical quantities are strictly real values: $Q = 0$ and $\overbar{Z} = R$. The DC nodes are power-controllable $P_{dc}$-nodes or voltage-controllable $V_{dc}$-nodes  (see Table \ref{NodeTypes}). 

The ICs can operate in different operation modes. Generally, two quantities are controlled simultaneously due to the decoupling of the $d$ and $q-$axes of the dq reference frame in the converter's inner control loops. 
Therefore, in grid-following mode, typically the DC voltage and reactive power are regulated in the IC nodes $(k,l) \in \mathcal{L}_{V_{dc}Q}$ or the active and reactive powers $(k,l) \in \mathcal{L}_{PQ}$ (see Table \ref{NodeTypes}). 
When the IC operates in grid-forming mode, the AC voltage magnitude, angle, and frequency are regulated at the IC nodes $(k,l) \in \mathcal{L}_{VV}$. The AC and DC nodes of the grid-forming ICs cannot be modelled using the traditional load flow theory. Indeed, while the AC side can be treated as a conventional slack node, the DC side cannot, as no quantity is regulated (neither $P_{dc}$ nor $E_{dc}$). \footnote{The same occurs for the grid-following ICs that regulate the DC voltage.}. 
Therefore, the load flow model proposed in reference \cite{willem_LF} has to be extended to account for grid-following ICs. 
\renewcommand{\arraystretch}{1.5}
\begin{table}[ht]
\caption{Different types of nodes in hybrid AC/DC networks and their known and unknown variables.}
\resizebox{1.05\columnwidth}{!}{%
\begin{tabular}{lllll}
\textbf{Bus Type   }             & \textbf{Ctrl. mode }                     & \textbf{Known var. }                                             & \textbf{Unknown var. } & \textbf{Index}\\ \hline
AC slack                   &                                  & $\lvert E_{ac} \rvert$, $\angle E_{ac}$                               & $P_{ac}$,$Q_{ac}$       & $s \in \mathcal{N}_{slack}$   \\ \hline
$P_{ac}$, $Q_{ac}$                 &                                  & $P_{ac}$,$Q_{ac}$                                                & $\lvert E_{ac} \rvert$, $\angle E_{ac}$    & $i \in \mathcal{N}_{PQ}$            \\ \hline
$P_{ac}$, $\lvert V_{ac} \rvert$                 &                                  & $P_{ac}$,$\lvert E_{ac} \rvert$                                                & $Q_{ac} $, $\angle E_{ac}$  & $i \in \mathcal{N}_{PV}$              \\ \hline
\multirow{2}{*}{$IC_{ac}$} & \textit{Power}                & $P_{ac}$ $Q_{ac}$     & $\lvert E_{ac} \rvert$, $\angle E_{ac}$           &  $l \in \mathcal{L}_{PQ}$  \\ \cdashline{2-5}
                        & \textit{Voltage}                    & $Q_{ac}$              & $P_{ac}$ $\lvert E_{ac} \rvert$ $\angle E_{ac}$     &  $l \in \mathcal{L}_{V_{dc}Q}$        \\ \cdashline{2-5} 
                        & \textit{Grid-forming} & $\lvert E_{ac} \rvert$, $\angle E_{ac} $                       & $P_{ac}$, $Q_{ac}$  &   $l \in \mathcal{L}_{VV}$            \\ \hline
\multirow{2}{*}{$IC_{dc}$} & \textit{Power}                & $P_{dc}$              & $E_{dc}$           &  $k \in \mathcal{L}_{PQ}$   \\ \cdashline{2-5}
                        & \textit{Voltage}                    & $E_{dc}$              & $P_{dc}$           &   $k \in \mathcal{L}_{V_{dc}Q}$  \\  \cdashline{2-5} 
                        & \textit{Grid-forming} & -                       & $P_{dc}$, $E_{dc}$  &   $k \in \mathcal{L}_{VV}$            \\ \hline
$P_{dc}$                     &                                  & $P_{dc}$                                                    & $E_{dc}$          & $j \in \mathcal{M}_{P}$    \\ \hline
$V_{dc}$                     &                                  & $E_{dc}$                                                    & $P_{dc}$           & $j \in \mathcal{M}_{V}$   \\ \hline
\end{tabular} \label{NodeTypes}
}
\end{table}
\normalsize

\subsubsection{Load flow model}

The generic three-phase load flow equations for a hybrid AC/DC network, proposed in \cite{willem_LF}, are presented in \eqref{eq:PFmodel}. 
\begin{subequations}
\small
\allowdisplaybreaks
\begin{align}
&\text{\textbf{AC nodes}} \nonumber \\
        &\Re \Big\{ \overbar{E}_{i}^{\phi} \sum\nolimits_{n \in \mathcal{N}} \underbar{Y}_{i,n}^{ac} \underbar{E}_{n}^{\phi} \Big\} = P^{\phi \ast}_{i} , &&  \forall i \in \mathcal{N}_{PQ} \cup \mathcal{N}_{PV} 
        \label{PQ_p}\\ 
        &\Im \Big\{ \overbar{E}_{i}^{\phi} \sum\nolimits_{n \in \mathcal{N}} \underbar Y_{i,n}^{ac} \underbar E_{n}^{\phi} \Big\} = Q^{\phi \ast}_{i} , &&  \forall i \in \mathcal{N}_{PQ} \label{PQ_q}\\ 
        & \big( {E_{i}^{\phi \prime}} \big) ^2 + \big( {E_{i}^{\phi \prime \prime}} \big)^2 = \lvert {E_{i}^{\phi \ast}} \rvert ^2 , 
        &&  \forall i \in \mathcal{N}_{PV} \label{PV} \\
        \vspace{5pt}
&\text{\textbf{DC nodes}}  \nonumber \\
        & E_{j} \sum\nolimits_{m \in \mathcal{M}} Y_{j,m}^{dc} E_{m}  = P^{\ast}_{j} , 
        && \forall j \in \mathcal{M}_{P} \label{P_dc}\\ 
        & E_{j} = E^{\ast}_{j} , 
        && \forall j \in \mathcal{M}_{V} \label{V_dc} \\
        \vspace{-5pt}
&\text{\textbf{IC nodes}} \nonumber \\ 
       & E_{k}^{\ast} \Big( Y_{(k,k)}^{dc} E_{k}^{\ast} + \sum\nolimits_{\substack{m \in \mathcal{M} \\
              m \neq k}} Y_{(k,m)}^{dc} E_{m} \Big)  
              && \nonumber \\
       &  \qquad \quad  = \Re  \Big\{ \overbar{E}_{l}^+ \sum\nolimits_{n \in \mathcal{N}} \underbar Y_{(l,n)}^{ac} \underbar E_{n}^+ \Big\},  
       && \forall (l,k) \in \mathcal{L}_{V_{dc}Q}   \label{Edc}\\
        &  \Re \Big\{ \overbar{E}_{l}^+ \sum\nolimits_{n \in \mathcal{N}} \underbar Y_{(l,n)}^{+ ac} \underbar E_{n}^+ \Big\}   = P^{\ast}_l  ,  
        && \forall l \in \mathcal{L}_{PQ} 
        \label{EP} \\ 
       &  \Im \Big\{ \overbar{E}_{l}^+ \sum\nolimits_{n \in \mathcal{N}} \underbar Y_{(l,n)}^{+, ac} \underbar E_{n}^+ \Big\}   = Q^{\ast}_l ,  &&  \forall l \in \mathcal{L}_{PQ} \cup \mathcal{L}_{V_{dc}Q} 
       \label{EQ} \\
        &E^{0  \prime}_l  = 0 , \qquad \ E^{0  \prime \prime}_l  = 0 , && \forall l \in \mathcal{L}_{PQ} \cup \mathcal{L}_{V_{dc}Q} 
        \label{E0} \\ 
        &E^{-  \prime}_l  = 0 , \qquad E^{-  \prime \prime}_l  = 0,  && \forall l \in \mathcal{L}_{PQ} \cup \mathcal{L}_{V_{dc}Q} 
        \label{E-} 
\end{align} \label{eq:PFmodel}
\end{subequations}
\vspace{-2pt}
The asterisk $ \scriptstyle{\square}^{\ast}$ refers to the controllable (i.e. fixed) variables, $\phi$ is the phase: $\phi \in \{a,b,c \}$ and the symmetrical components are indicated by $\{ +,-,0\}$. The underbar $\underbar{$\scriptstyle{\square}$}$ indicates the complex conjugate, and the prime symbols $ \scriptstyle{\square}^{\prime}$ and $ \scriptstyle{\square}^{\prime \prime}$ refer to the real and imaginary parts of the complex phasors. 

As previously mentioned, the model presented in \cite{willem_LF} is limited and cannot incorporate the grid-forming operation of the ICs, as only the AC-side variables of the grid-forming IC are controllable. 
The DC-side quantities, however, can be computed from the active power balance of the IC \eqref{Pblanced}
\footnote{The grid-forming IC model is only derived for the single-phase equivalent but can be computed for any generic three-phase unbalanced system.}.
\vspace{-2pt}
\begin{flalign}
    P_{l} + P^{loss}_{(l,k)}  = P_{k}, \qquad\forall (l,k) \in \mathcal{L}_{VV} 
    \label{Pblanced}
\end{flalign}
\vspace{-2pt}
\noindent Therefore, using \eqref{PQ_p} and \eqref{P_dc}, the power balance in \eqref{Pblanced} is reformulated.
\vspace{-2pt}
\begin{flalign} \label{eq:start_bal}
    & \Re \Big\{ \overbar{E}_{l} \sum_{n \in \mathcal{N}} \underbar Y_{(l,n)}^{ac} \underbar E_{n} \Big\}  +  P^{loss}_{(l,k)} = E_{k} \sum_{m \in \mathcal{M}} Y_{(k,m)}^{dc} E_{m} 
\end{flalign}
\vspace{-2pt}
The controllable variable of the grid-forming IC is the AC voltage phasor $\overbar{E}_l^{\ast}$; its magnitude and angle $( \lvert \overbar{E}_l \rvert \angle \theta_l)$ or its real and imaginary parts $(E_l^{\prime} + j E_l^{\prime \prime})$. In this work, the problem is formulated in rectangular coordinates\footnote{The power flow model can also be derived in polar coordinates.}. The controllable variable $\overbar{E}_{l}^{\ast}$ is isolated from  \eqref{eq:start_bal}:
\vspace{-2pt}
\begin{flalign}
    &\Re \Big\{ \overbar{E}_{l}^{\ast} \sum\nolimits_{\substack{n \in \mathcal{N}\\ n \neq l}}
                  \underbar Y_{(l,n)}^{ac} \underbar E_{n} \Big\} + \Re \Big\{ \overbar{E}_{l}^{\ast} \underbar Y_{(l,l)}^{ac} \underbar E_{l}^{\ast} \Big\}  +  P^{loss}_{(l,k)}  \nonumber \\
    & - E_{k} \sum\nolimits_{m \in \mathcal{M}} Y_{(k,m)}^{dc} E_{m} = 0, \hspace{20pt} \forall (l,k) \in \mathcal{L}_{VV}  \label{eq:VV_1}
\end{flalign}
\vspace{-2pt}
To simplify the derivation, we assume the IC buses are connected to the AC and DC grids through a single connection\footnote{When multiple lines are connected to the IC bus, expressions in \eqref{eq:VV_2} must be suitably modified.}. Furthermore, the complex voltage phasors are reformulated in Cartesian coordinates. Therefore, \eqref{eq:VV_1} becomes a quadratic expression in the controllable variable $E_l^{ \prime \ast }$:
\vspace{-2pt}
\begin{flalign}
    & \Big[ G_{(l,l)} \Big] {E_l^{ \prime \ast }}^2 + \Big[ G_{(l,i)} E_i^{\prime} -  B_{(l,i)} E_i^{\prime \prime} \Big]  E_l^{ \prime \ast } + \Big[ P^{loss}_{(l,k)}  + G_{(l,l)}  {E_l^{\prime \prime}}^2  \nonumber\\
    & + G_{(l,i)} E_i^{\prime \prime} E_l^{\prime \prime} + B_{(l,i)} E_i^{\prime} E_l^{\prime \prime} - G_{k,k} E_k^2 - G_{k,j} E_k E_j \Big] = 0  \nonumber \\
    & \hspace{150pt} \forall (l,k) \in \mathcal{L}_{VV}  
    \label{eq:VV_2} 
\end{flalign}
\vspace{-2pt}
The quadratic expression in \eqref{eq:VV_2} has two solutions: a technically feasible one close to 1 p.u. and an infeasible one around 0 p.u. The load flow model for the grid-forming IC becomes: 
\vspace{-2pt}
\begin{flalign}
    & \begin{cases}
        & E_l^{\prime \ast } = \dfrac{-b}{2a} + \dfrac{\sqrt{b^2 - 4ac}}{2a}  \\
        & E_l^{\prime \prime \ast } = 0  \hspace{110pt} \forall (l,k) \in \mathcal{L}_{VV} \label{eq:VVextra_1}
    \end{cases} \\
    &  \text{with} \nonumber\\
    &  \qquad a = G_{(l,l)}  \nonumber \\
    &  \qquad b = G_{(l,i)} E_i^{\prime} -  B_{(l,i)} E_i^{\prime \prime}   \nonumber \\
    &  \qquad c = G_{(l,l)}  {E_l^{\prime \prime}}^2 + G_{(l,i)} E_i^{\prime \prime} E_l^{\prime \prime} + B_{(l,i)} E_i^{\prime} E_l^{\prime \prime}  \nonumber  \\
    & \qquad \hspace{12pt} - G_{k,k} E_k^2 - G_{k,j} E_k E_j +  P^{loss}_{(l,k)}   \nonumber
\end{flalign}
\vspace{-2pt}
Typically, the reference angle of the voltage at the slack node is set to zero. Therefore, in a Cartesian coordinate system, the imaginary part of the voltage becomes zero $E_l^{\prime \prime \ast} = 0$.

\vspace{3pt}
\subsubsection{Interfacing converters loss model}
The IC losses, denoted by $P^{loss}_{(l,k)}$, can be included to improve the accuracy of the load flow model \eqref{eq:PFmodel}. As presented in \cite{Graovac2009, EU_converter_losses}, the total IC losses can be modelled as the sum of the conduction and switching losses\footnote{The blocking losses are typically neglected but can be included in the application of series stacked devices \cite{EU_converter_losses}.}. The losses are assumed to be independent of the operational mode of the IC.

The conduction losses refer to the ohmic losses in the transistor and the diode. These losses are characterised in the component's datasheet: for the transistor, the collector-emitter voltage drop depends on the collector current, while for the diode, the forward voltage drop is determined by the forward current.
Typically, these parameters match, and the conduction losses can be approximated as a constant voltage $V_0$ and an equivalent series resistance $R_{eq}$\footnote{In case the IGBT and diode characteristics do not match, the equivalent parameters can be formulated using the average duty cycle of the switching pulse}.

The switching losses are attributed to the transistor's turn-on and turn-off energy loss and the diode's reverse recovery loss. The components' datasheet typically provides these switching energies as average values under nominal test conditions. This work approximates these energy losses by a second-order polynomial proportional to the DC voltage.
\vspace{-2pt}
\begin{flalign}
E_{on} + E_{off} + E_{rec} = \frac{E_k}{E_{nom}} \left( u +  v \lvert \overbar{I}_{l} \rvert + w \lvert \overbar{I}_{l} \rvert ^2 \right) \label{eq:losses_start}
\end{flalign}
\vspace{-2pt}
The parameters $E_{nom}, \ u, \ v \, \text{and} \ w$ can be obtained from the semiconductor datasheet or through a system identification algorithm.
The associated switching power losses are computed as the product of these switching energies and the switching frequency $f_{sw}$.
\vspace{-2pt}
\begin{flalign}
    P^{loss}_{(l,k)} &= P^{cond}_{(l,k)} + P^{switching}_{(l,k)} \label{eq:losses} \\
    & = \frac{2 \sqrt{2} V_{0}}{\pi} \lvert \overbar{I}_l \rvert  + R_{0} \lvert \overbar{I}_l \rvert^2 + \frac{f_{sw}  E_k}{E_{nom}} \left( u +  v \lvert \overbar{I}_{l} \rvert + w \lvert \overbar{I}_{l} \rvert ^2 \right)  \nonumber 
\end{flalign}
\vspace{-2pt}
To integrate the losses model into the load flow model, expression \eqref{eq:losses} is reformulated as a function of the state variables, namely the AC and DC nodal voltage phasors.
\vspace{3pt}
\subsubsection{Interfacing converter filter model}
The IC filter is not explicitly included in the power flow model. This contrasts with the original load flow model proposed in \cite{willem_LF}, where the active and reactive power losses in the filter are included as an additional term. 
In this work, the filter is included in the admittance matrix as an equivalent branch (see Figure \ref{fig:gengrid}). It is known that any filter consisting of capacitive and inductive elements, since being a passive and reciprocal two-port model, can be modelled as a $\Pi$-equivalent branch and integrated into the system's admittance matrix.
This approach greatly simplifies the load flow model and, therefore, the sensitivity coefficient-based OPF model.
\vspace{-5pt}
\subsection{Analytical sensitivity coefficient model}
The OPF problem requires a grid model to account for the grid constraints, such as voltage and ampacity limits. This work uses a linear first-order Taylor approximation of the previously presented load flow model, referred to as sensitivity coefficients. These SCs are the partial derivatives of the nodal voltages and branch current flows with respect to the nodal power injections. Therefore, instead of using the non-linear, non-convex load flow equations, the nodal voltages and branch current constraints can be linearly formulated using the SCs, as given in \eqref{sc_voltage} and \eqref{sc_current}.
\vspace{-2pt}
\begin{eqnarray}
    &\lvert \overbar{\mathbf{E}} \rvert^{t} = \lvert \overbar{\mathbf{E}} \rvert^{t-1}  + \left[ \mathbf{\frac{\partial \lvert \overbar{E} \rvert }{\partial P}} \right]^{t} \Delta {\mathbf{P}}^{t} + \left[\mathbf{\frac{\partial \lvert \overbar{E} \rvert }{\partial Q}}\right]^{t} \Delta {\mathbf{Q}}^{t}, \label{sc_voltage} \\
    &\lvert \overbar{\mathbf{I}} \rvert^{t} = \lvert \overbar{\mathbf{I}} \rvert^{t-1} + \left[\mathbf{\frac{\partial \lvert \overbar{I} \rvert }{\partial P}}\right]^{t} \Delta {\mathbf{P}}^{t} + \left[\mathbf{\frac{\partial \lvert \overbar{I} \rvert }{\partial Q}}\right]^{t} \Delta {\mathbf{Q}}^{t},  \label{sc_current}
\end{eqnarray}
\vspace{-2pt}
where $\lvert \overbar{\mathbf{E}} \rvert^{t}$ and $\lvert \overbar{\mathbf{I}} \rvert^{t}$ are the vectors of the nodal voltages and current flows (AC and DC) at timestep $t$. $\Delta \mathbf{P}^{t}$ $\Delta \mathbf{Q}^{t}$ are the vectors of the change in power injections between timesteps $t$ and $t-1$. The SCs are expressed in matrix form.


The linearized approach allows the OPF problem to be solved recursively, with time resolution ranging from seconds to sub-seconds, making it particularly suitable for time-critical real-time control applications
Furthermore, as presented below, the analytical computation of the SCs requires only the network's admittance matrix and the current state of the grid, which is provided by a state estimation process  \cite{christakou2015real}. It is demonstrated in \cite{gupta2019performance} that when the SCs are updated dynamically, i.e. at each timestep, the linear grid model is exact with respect to the load flow. 

The analytical model that allows for the analytical computation of the SCs has been originally presented in \cite{christakou2015real}. It solves a linear system: $\mathbf{A} \mathbf{u} = \mathbf{b}$, where $\mathbf{u}$ is the unknown vector of the SCs.
This method has been extended in \cite{willem_SC} for hybrid AC/DC networks and includes the ICs operating in grid-following mode (voltage and power mode). However, this extension does not include a grid-forming IC model or the detailed IC losses model presented above. 
Therefore, this section extends the analytical model of \cite{willem_SC} to include grid-forming ICs and a detailed IC losses model\footnote{Only the relevant extension is given here. The full development is presented in detail in \cite{willem_SC}.}.

To maintain generality, the SCs are computed with respect to $x$, where $x \in \mathcal{X}$, the set of all controllable variables.
\vspace{-2pt}
\begin{align}
&\mathcal{X} = \left\{ P^{\phi \ast}_{i},Q^{\phi \ast}_{i}, \lvert \overbar{E}_{i}^{\phi \ast} \rvert ,P^{\ast}_{j},E^{\ast}_{j},P^{\ast}_l,Q^{\ast}_l , \lvert \overbar{E}_l^{\ast} \rvert, \angle \overbar{E}_l^{\ast}, E_{k}^{\ast} \right\} \nonumber \\ 
& \qquad \qquad \forall \ i \in \mathcal{N}, \forall \ j \in \mathcal{M}, \forall \ (l,k) \in \mathcal{L}  \label{eq:X}
\end{align}
\vspace{-2pt}
The vector $\mathbf{u}(x)$ of the unknown voltage SCs with respect to $x$ is given as:
\vspace{-2pt}
\begin{flalign}
 &\mathbf{u}(x) =  \left[ \frac{\partial \overbar{E}_i^{\prime}}{ \partial x}, \frac{\partial \overbar{E}_i^{\prime \prime}}{ \partial x},
                                    \frac{\partial {E}_j}{ \partial x} , 
                                    \frac{ \partial \overbar{E}_l^{\prime}}{ \partial x}, \frac{ \partial \overbar{E}_l^{\prime \prime}}{ \partial x} ,
                                    \frac{ \partial {E}_k}{ \partial x}
                            \right], \nonumber \\ 
&  \qquad  \qquad  \forall \ i \in \mathcal{N},  \ j \in \mathcal{M},  \ (l,k) \in \mathcal{L},  \ x \in \mathcal{X}. \label{d_dX}
\end{flalign}
\vspace{-2pt}
Therefore, the linear system  $\mathbf{A} \mathbf{u}(x) = \mathbf{b}(x)$ is solved for each $x$ to calculate the partial derivatives of all voltages (AC and DC) with respect to every controllable variable.

Trivially, the nodal voltages of the voltage-controllable nodes are fixed and cannot be influenced by any control variable other than themselves. Therefore, the partial derivative of any nodal voltage with respect to itself is 1 \eqref{eq:sc_1_1}, and the SCs of these voltages with respect to any other controllable variable are zero  \eqref{eq:sc_1_2},\eqref{eq:sc_1_3}. 
\vspace{-2pt}
\begin{subequations}
\begin{align}
   &\frac{\partial \lvert \overbar{E}_l^{\ast} \rvert }{ \partial \lvert \overbar{E}_l^{ \ast} \rvert }  = \frac{\partial \angle \overbar{E}_l^{\ast} }{ \partial \angle \overbar{E}_l^{ \ast}  } = 1, && \forall l \in \mathcal{L}_{VV}, \label{eq:sc_1_1} \\
   &\frac{\partial \lvert \overbar{E}_l^{\ast} \rvert }{ \partial x }  = 0,&& \forall l \in \mathcal{L}_{VV}, \ \forall x \in \mathcal{X} \setminus \{ \lvert \overbar{E}_{l}^{\ast} \rvert \}  \label{eq:sc_1_2} \\
   &\frac{\partial \angle \overbar{E}_l^{\ast}  }{ \partial x }  = 0,&& \forall l \in \mathcal{L}_{VV}, \ \forall x \in \mathcal{X} \setminus \{ \angle \overbar{E}_{l}^{\ast} \}  \label{eq:sc_1_3} \vspace{-2pt}
    \end{align} \label{eq:sc_1}
\end{subequations}    
Consequently, the only SCs of the nodal voltage at the DC terminals of the grid-forming ICs, represented as $\frac{\partial E_k}{\partial x}, \forall k \in \mathcal{L}_{VV}$ need to be computed.
Their analytical expression is obtained by taking the partial derivative of the corresponding load flow equation \eqref{eq:VV_2}:
\vspace{-2pt}
\begin{flalign}
    & 2 \Big[ G_{(l,l)} \Big] {\overbar{E}_l^{\ast \prime}} \frac{\partial E_l^{\ast \prime}}{\partial x} +  \Big[ G_{(l,i)} \overbar{E}_i^{\prime} -  B_{(l,i)} E_i^{\prime \prime} \Big]   \frac{\partial \overbar{E}_l^{\ast \prime}}{\partial x} + \nonumber \\
    & \Big[ G_{(l,i)} \frac{\partial \overbar{E}_i^{\prime}}{\partial x}  -  B_{(l,i)} \frac{\partial \overbar{E}_i^{\prime \prime }}{\partial x}  \Big]  \overbar{E}_l^{\ast \prime} + \nonumber \\
    & \Big[  2 G_{(l,l)}  {\overbar{E}_l^{\ast\prime \prime}} \frac{\partial \overbar{E}_l^{\ast\prime \prime}}{\partial x} +  G_{(l,i)} \frac{\partial \overbar{E}_i^{\prime \prime}}{\partial x} \frac{\partial \lvert \overbar{E} \rvert}{\partial x} _l^{\ast\prime \prime} + G_{(l,i)} \overbar{E}_i^{\prime \prime} \frac{\partial 
      \overbar{E}_l^{\ast\prime \prime}}{\partial x} + \nonumber \\
    & B_{(l,i)} \frac{\partial E_i^{\prime}}{\partial x}  E_l^{\prime \prime} + B_{(l,i)} \overbar{E}_i^{\prime} \frac{\partial \overbar{E}_l^{\ast\prime \prime}}{\partial x}  - 
      2 G_{k,k} E_k^2 \frac{\partial E_k}{\partial x}  - \nonumber \\
    & G_{k,j} \frac{\partial E_k}{\partial x}  E_j - G_{k,j} E_k \frac{\partial E_j}{\partial x}  +  \frac{\partial P^{loss}_{(l,k)} }{\partial x}  \Big] = 0 \nonumber \\
    & \hspace{150pt} \forall (l,k) \in \mathcal{L}_{VV}  \label{eq:sc_2} \vspace{-2pt}
\end{flalign}
\vspace{-4pt}
\noindent By rearranging the terms, we obtain the following expression:
\vspace{-3pt}
\begin{flalign}
    & -\Big[ 2 G_{(l,l)} {\overbar{E}_l^{\ast \prime}} + G_{(l,i)} \overbar{E}_i^{\prime} -  B_{(l,i)} \overbar{E}_i^{\prime \prime} \Big]   \frac{\partial \overbar{E}_l^{\ast \prime}}{\partial x}  \nonumber \\
    & -\Big[ 2 G_{(l,l)} {\overbar{E}_l^{\ast \prime \prime}} 
      +G_{(l,i)} \overbar{E}_i^{\prime \prime}  
      +B_{(l,i)} \overbar{E}_i^{\prime} \Big]  \frac{\partial \overbar{E}_l^{\ast \prime \prime}}{\partial x} = \nonumber  \\
    & \hspace{12pt} \Big[ G_{(l,i)} \overbar{E}_l^{\ast \prime} 
      +B_{(l,i)}  \overbar{E}_l^{\ast \prime \prime} \Big] \frac{\partial \overbar{E}_i^{\prime}}{\partial x}  \nonumber  \\ 
    & +\Big[-B_{(l,i)} \overbar{E}_l^{\ast \prime}
      +G_{(l,i)} \overbar{E}_l^{\ast \prime \prime} \Big] \frac{\partial \overbar{E}_i^{\prime \prime}}{\partial x} \nonumber  \\
    &  -\Big[ 2 G_{k,k} E_k^2 + G_{k,j} E_j \Big] \frac{\partial E_k}{\partial x}  
      - \Big[G_{k,j} E_k \Big] \frac{\partial E_j}{\partial x}  
      +\frac{\partial P^{loss}_{(l,k)} }{\partial x}  \nonumber \\
    & \hspace{150pt} \forall (l,k) \in \mathcal{L}_{VV}  \label{eq:sc_3} \vspace{-4pt}
\end{flalign}
This expression represents a linear combination of the unknown voltage SCs in \eqref{d_dX}, with coefficients dependent solely on the grid state and the admittance matrix. Typically, the voltage magnitude and angle of the grid-forming IC are controlled rather than their real and imaginary components. Using the identities provided in \eqref{SCidentity2}, \eqref{eq:sc_3} is reformulated in terms of the controllable variables $\lvert \overbar{E}_l \rvert^{\ast}$ and $\angle \overbar{E}_l^{\ast}$.
\vspace{-2pt}
\begin{flalign}
        & \frac{\partial \overbar{E}^{\prime}}{ \partial x} = \frac{ \overbar{E}^{\prime}}{\lvert \overbar{E} \rvert} \frac{\partial \lvert \overbar{E} \rvert}{\partial x} - \overbar{E}^{\prime \prime} \frac{\partial \angle \overbar{E} }{\partial x}  \nonumber \\
        & \frac{\partial \overbar{E}^{\prime \prime}}{ \partial x} = \frac{ \overbar{E}^{\prime \prime}}{\lvert \overbar{E} \rvert} \frac{\partial \lvert \overbar{E} \rvert}{\partial x} + \overbar{E}^{\prime} \frac{\partial \angle \overbar{E} }{\partial x}   \label{SCidentity2}
\end{flalign}
\vspace{-2pt}
\begin{flalign}
    & \hspace{12pt} \Big[ -\big( 2 G_{(l,l)} {E_l^{\ast \prime}} + G_{(l,i)} E_i^{\prime} -  B_{(l,i)} E_i^{\prime \prime} \big)  \frac{ \overbar{E}_l^{\prime}}{\lvert \overbar{E}_l \rvert} \nonumber \\
    &  \hspace{30pt}  -\big( 2 G_{(l,l)} {E_l^{\ast \prime \prime}} +G_{(l,i)} E_i^{\prime \prime}  +B_{(l,i)} E_i^{\prime} \big)   \frac{ \overbar{E}_l^{\prime \prime}}{\lvert \overbar{E}_l \rvert} \Big] \frac{\partial \lvert \overbar{E}_l \rvert^{\ast}}{\partial x} \nonumber \\
    & + \Big[ \big( 2 G_{(l,l)} {E_l^{\ast \prime}} + G_{(l,i)} E_i^{\prime} -  B_{(l,i)} E_i^{\prime \prime} \big)   \overbar{E}_l^{\prime} \nonumber  \\
    & \hspace{30pt}  -\big( 2 G_{(l,l)} {E_l^{\ast \prime \prime}} +G_{(l,i)} E_i^{\prime \prime}  +B_{(l,i)} E_i^{\prime} \big)   \overbar{E}_l^{\prime \prime} \Big] \frac{\partial \angle \overbar{E}_l^{\ast} }{\partial x}   = \nonumber  \\
      & \hspace{12pt} \Big[ G_{(l,i)} E_l^{\ast \prime} 
      +B_{(l,i)}  E_l^{\prime \prime} \Big] \frac{\partial E_i^{\prime}}{\partial x}  \nonumber  \\ 
    & +\Big[-B_{(l,i)} E_l^{\ast \prime}
      +G_{(l,i)} E_l^{\prime \prime} \Big] \frac{\partial E_i^{\prime \prime}}{\partial x} \nonumber  \\
    &  -\Big[ 2 G_{k,k} E_k^2 + G_{k,j} E_j \Big] \frac{\partial E_k}{\partial x}  
      -\Big[G_{k,j} E_k \Big] \frac{\partial E_j}{\partial x}  
      +\frac{\partial P^{loss}_{(l,k)} }{\partial x}  \nonumber \\
    & \hspace{150pt} \forall (l,k) \in \mathcal{L}_{VV}  \label{eq:sc_4}
\end{flalign}
The partial derivates $\frac{\partial \lvert \overbar{E}_l \rvert^{\ast}}{\partial x}$ and $\frac{\partial \angle \overbar{E}_l^{\ast} }{\partial x} $ in the left-hand side of the expression can be substituted by \eqref{eq:sc_1}. 
This expression is linear in the unknown SCs $\mathbf{u}(x)$ and is included in the linear system of equations $\mathbf{A} \mathbf{u}(x) = \mathbf{b}(x) $. This system is solved for each $x \in \mathcal{X}$ to compute the voltage sensitivity coefficient matrix\footnote{The partial derivative of the loss term $P_{(l,k)}^{loss}$ with respect to $x$ is not presented due to space limitations. 
}. 
Once the voltage SCs are computed, the current flow SCs can be calculated using the network's admittance matrix \cite{willem_SC}. The SCs are used to define the grid constraints within the OPF problem, as demonstrated in \eqref{sc_voltage} \eqref{sc_current}.

\vspace{-5pt}

\vspace{-5pt}
\section{Optimal control architecture}
\label{sec:architecture}

\subsection{State machine}

\begin{figure}[H] \vspace{-7pt}
    \centering
    \includegraphics[width=1\linewidth]{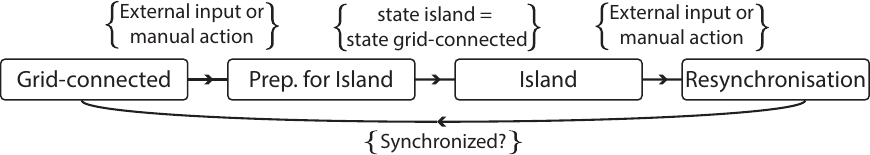}
    \caption{The state machine architecture. The state transition conditions are indicated by the curly braces brackets}
    \label{fig:SM} \vspace{-10pt}
\end{figure}

The backbone of the optimal control architecture is the state machine, which ensures correct transitions between the different operating states (see Figure \eqref{fig:SM}). The four operating states are \textit{Grid-connected}, \textit{Prepare-for-island}, \textit{Island}, and \textit{Resynchronisation}.
The state machine also facilitates communication between the various resources, the optimal power flow computation, and the synchro-check relay, which is responsible for opening and closing the main breaker\footnote{We assume that the AC/DC microgrid is connected to the upstream network at a single point, the Grid Connection Point (GCP), which is equipped with a controllable main breaker.} (see Figure \ref{fig:gengrid}). 

\vspace{-10pt}
\subsection{Optimal power flow model}
The general optimal power flow model is presented in \eqref{eq:opf}. The OPF problem is solved at every timestep $t$.
Depending on the operating state, the objective function is adapted, or additional constraints are introduced to ensure a smooth transition and prevent voltage surges or inrush currents.

The \textbf{objective function} of the general OPF problem, presented in \eqref{eq:opf_obj}, aims to minimise multiple objectives: 1) the reactive power at the GCP (see Figure \ref{fig:gengrid}), 2) the losses in both the AC and DC grids, 3) the active and reactive power circulated by the ICs, 4) the voltage deviation in the DC grid, and 5) the deviation of the SoC from its reference value. This applies to the SoC of the ESS, which ensures that the power balance is satisfied during islanded operation. Typically, an ESS is used in practice as it can provide bidirectional power control.
Each objective term is scaled by a weight factor $w_{1...6}$, calculated by normalising the terms in the objective function and assigning a weight depending on their relative importance. Most elements of the objective function are largely independent and have minimal influence on one another.

The \textbf{constraints} of the general optimisation problem are presented in \eqref{eq:opf_v} - \eqref{eq:opf_form}. 
Equations \eqref{eq:opf_v}, \eqref{eq:opf_i}, \eqref{eq:opf_ploss} and \eqref{eq:opf_qloss} define the SC-based linearised grid model for the nodal phase-to-ground voltage magnitudes, branch current flow, and active and reactive grid losses\footnote{Similar to the current SCs, the SCs of the line losses are calculated using the voltage SCs and the network's admittance matrix.}. 
The nodal voltage and branch ampacity limits are specified in \eqref{eq:opf_vlim} and \eqref{eq:opf_ilim}. 
Equations \eqref{eq:opf_pi}, \eqref{eq:opf_qi}, \eqref{eq:opf_pl}, \eqref{eq:opf_ql}, and \eqref{eq:opf_pj} present the capacity limits of the DERs and the ICs.
The ESS's SoC is modelled in \eqref{eq:opf_SOCmodel}, and its safe operational range is defined in \eqref{eq:opf_soclim}. Here, $\eta$ denotes the efficiency of the storage device, $T_s$ is the timestep of the control framework, and $E^{tot}$ denotes the ESS's total usable energy capacity. 
Equation \eqref{eq:opf_SOCmodel} models how the SoC evolves when the ESS is charged (or discharged) with a power $P_{ess}$.
The active and reactive power of the slack node is defined in equations \eqref{eq:opf_sq} and \eqref{eq:opf_sp}. These powers are not controllable, unlike, e.g. \textit{PQ nodes}, and are modelled explicitly using the power balance of the entire hybrid AC/DC grid.
Equation \eqref{eq:opf_ramp} introduces the ramp constraint on the active power of the IC that will transition to grid-forming operation. This guarantees a smoother profile immediately after the network is islanded or resynchronised. 
The constraint in \eqref{eq:opf_pic} specifies the mapping between the active power on the AC side and the DC side of the ICs using the power losses $P_{(l,k)}^{loss}$. While the SCs intrinsically encapsulate this relationship for the load flow constraints \eqref{eq:opf_v}-\eqref{eq:opf_qloss}, this mapping must be explicitly formulated when we want to constrain the IC's AC and DC power injections.
Equation \eqref{eq:opf_form} represents the power balance in the DC grid for the voltage-controllable IC. This IC will transition to grid-forming mode. The additional expression is necessary because the active power is not a controllable variable in the DC voltage control mode (only $V_{dc} \text{ and } Q$ are controllable). 
%
\vspace{-7pt}
\begin{mini!}|l|[3] 
{\begin{aligned} & P_{i}^t, Q_{i}^t, P_{l}^t, Q_{l}^t, \\ & P_{k}^t, E_k^t, E_j^t, \lvert \overbar{E}_l^t \rvert \end{aligned}}
{\begin{aligned}
    & w_1 \lvert \lvert Q_{gcp}^t \rvert \rvert  + w_2\lvert \lvert P_{losses}^{t} \rvert \rvert + w_3\lvert \lvert P_{IC}^t \rvert \rvert + \\
    & w_4\lvert \lvert Q_{IC}^t \rvert \rvert + w_5 \lvert \lvert \ \lvert \overbar{E}^{ref} \rvert - \lvert \overbar{\mathbf{E}}_{j} \rvert^{t} \ \rvert \rvert + \\
    & w_6\lvert \lvert SoC_{ref} - SoC^t \rvert \rvert 
  \end{aligned} \label{eq:opf_obj} } 
{\label{eq:opf} }{}  
\addConstraint{  \Delta \lvert \overbar{\mathbf{E}} \rvert^{t} = \mathbf{K}_{P}^{E,t} \Delta {\mathbf{P}}^{t} + \mathbf{K}_{Q}^{E,t} \Delta {\mathbf{Q}}^{t} + \mathbf{K}_{E}^{E} \Delta \lvert \overbar{\mathbf{E}} \rvert^{t}  \label{eq:opf_v} }
\addConstraint{  \Delta \lvert \overbar{\mathbf{I}} \rvert^{t} = \mathbf{K}_{P}^{I,t} \Delta {\mathbf{P}}^{t} + \mathbf{K}_{Q}^{I,t} \Delta {\mathbf{Q}}^{t} + \mathbf{K}_{E}^{I,t} \Delta \lvert \overbar{\mathbf{E}} \rvert^{t} \label{eq:opf_i} }
\addConstraint{ \Delta {P}_{losses}^{t}  = \mathbf{K}_{P}^{P,t} \Delta \mathbf{P}^{t} + \mathbf{K}_{Q}^{P,t} \Delta \mathbf{Q}^{t} + \mathbf{K}_{E}^{P,t} \Delta \lvert \overbar{\mathbf{E}} \rvert^{t} \label{eq:opf_ploss}  }
\addConstraint{\Delta {Q}_{losses}^{t}  = \mathbf{K}_{P}^{Q,t} \Delta \mathbf{P}^{t} + \mathbf{K}_{Q}^{Q,t} \Delta \mathbf{Q}^{t} + \mathbf{K}_{E}^{Q,t} \Delta \lvert \overbar{\mathbf{E}} \rvert^{t} \label{eq:opf_qloss} }
\addConstraint{[ \mathbf{E}^{ac}_{min},  \mathbf{E}^{dc}_{min} ] \leq  \lvert \overbar{\mathbf{E}} \rvert^{t} \leq [ \mathbf{E}^{ac}_{max},  \mathbf{E}^{dc}_{max} ] \label{eq:opf_vlim} }
\addConstraint{\mathbf{0} \leq  \lvert \overbar{\mathbf{I}} \rvert^{t} \leq [ \mathbf{I}^{ac}_{max},  \mathbf{I}^{dc}_{max} ] \label{eq:opf_ilim} }
\addConstraint{-\widehat{P}_{i} \leq  P_{i}^t  \leq \widehat{P}_{i}, && \hspace{-141pt} \forall \ i \in \mathcal{N}_{PQ} \label{eq:opf_pi} }
\addConstraint{-\widehat{Q}_{i} \leq  Q_{i}^t  \leq \widehat{Q}_{i}, &&  \hspace{-141pt} \forall \ i \in \mathcal{N}_{PQ} \label{eq:opf_qi} }
\addConstraint{-\widehat{P}_{l} \leq  P_{l}^t  \leq \widehat{P}_{l}, && \hspace{-141pt} \forall \ l \in \mathcal{L}_{PQ} \label{eq:opf_pl} }
\addConstraint{-\widehat{Q}_{l} \leq  Q_{l}^t  \leq \widehat{Q}_{l}, &&  \hspace{-141pt} \forall \ l \in \mathcal{L}_{PQ} \cup \mathcal{L}_{V_{dc}Q} \label{eq:opf_ql} } 
\addConstraint{-\widehat{P}_{j} \leq  P_{j}^t  \leq \widehat{P}_{j}, && \hspace{-141pt} \forall \ j \in \mathcal{M}_{P} \label{eq:opf_pj} }
\addConstraint{ SoC^t = SoC^{t-1} - \eta P_{ess} \frac{T_s}{3600} \frac{1}{E^{tot}} \label{eq:opf_SOCmodel} }
\addConstraint{ SoC_{min} \leq  SoC^t  \leq SoC_{max}, \hspace{10pt} \label{eq:opf_soclim} }
\addConstraint{ Q_{s}^t = -\hspace{-10pt} \sum_{i \in \mathcal{N} \cap \mathcal{N}_{s}} \hspace{-8pt} Q_{i}^t -\sum_{l \in \mathcal{L}} Q_{l}^t+ Q_{losses}^{t} \label{eq:opf_sq} } 
\addConstraint{ P_{s}^t = -\hspace{-10pt} \sum_{i \in \mathcal{N} \cap \mathcal{N}_{s}} \hspace{-8pt} P_{i}^t -\sum_{l \in \mathcal{L}} P_{l}^t -\sum_{k \in \mathcal{L}} P_{k}^t - \sum_{j \in \mathcal{M}} P_{j}^t + P_{losses}^{t} \label{eq:opf_sp} }
\addConstraint{P_l^{t-1} - \Delta \widehat{P}< P_{l}^t < P_l^{t-1} + \Delta \widehat{P},  \hspace{24pt} \forall \ l \in \mathcal{L}_{V_{dc}Q} \label{eq:opf_ramp} }
\addConstraint{ P_{k}^t = -P_l^t - P^{loss}_{(l,k)}, \hspace{4pt} \forall (l,k) \in \mathcal{L}_{PQ} \cup \mathcal{L}_{V_{dc}Q} \cup \mathcal{L}_{VV} \label{eq:opf_pic} }
\addConstraint{P_{k}^t = -\sum_{j \in \mathcal{M}} P_{j}^t - \hspace{-15pt} \sum_{k \in \mathcal{L} \cap \mathcal{L}_{V_{dc}Q}} \hspace{-8pt} P_{k}^t + P_{\text{losses dc}}^{t},  \hspace{5pt}
\forall \ k \in \mathcal{L}_{V_{dc}Q}, \label{eq:opf_form} } 
\end{mini!} 
Depending on the operation state, the objective function and the constraints are modified:
\subsubsection{State 1: Grid connected}

The optimisation problem for the \textit{Grid-connected} mode is presented in \eqref{eq:opf}. It aims to operate the grid as efficiently as possible while ensuring that all grid and resource constraints are satisfied.
Additionally, it aims to bring the SoC of the ESS to the reference level, typically $50\%$, thereby ensuring maximum flexibility during islanded operation.
The third term in the objective function prevents the ICs from counteracting one another by minimising unwanted power circulation.

\subsubsection{State 2: Preparation for island}

The state machine transitions to \textit{Prepare for island} mode when triggered by an external signal, either manual or automatic (as shown in Figure \ref{fig:SM}).
The OPF aims to optimally control the set points of the DERs and the ICs, ensuring that the state of the grid remains identical after the islanding manoeuvre. This guarantees a seamless state transition. This is achieved by adding a term into the objective function that reduces the power flow through the GCP to zero\footnote{ During islanded operation, the power flow through the GCP is zero. This is enforced by a hard constraint defined by the topology of the islanded network. }. 
Furthermore, the active and reactive power processed by the IC, which is set to transition to grid-forming mode, is reduced to zero. The updated objective function is shown in \eqref{eq:opf_obj2}. The index $s$ denotes the AC slack node before islanding, while the index \textit{forming} specifies the IC that will transition to grid-forming operation.
\vspace{-2pt}
\begin{align}
    \eqref{eq:opf_obj} + w_7\lvert \lvert  P_{l,\text{forming}}^t \rvert \rvert + w_8\lvert \lvert  Q_{l,\text{forming}}^t \rvert \rvert + w_9\lvert \lvert  P_s^t \rvert \rvert  \label{eq:opf_obj2} 
    \vspace{-2pt}
\end{align}
\subsubsection{State 3: Island}
In \textit{Island} mode, the system's topology is fundamentally different. The \textit{Grid-connected} slack bus located at the GCP is now modelled as a zero-injection \textit{PQ} bus. The new slack node for the reactive power is situated on the AC side of the grid-forming IC. The slack node for the active power is in the DC grid at the ESS, which regulates the DC voltage. It is important to note that the grid-forming IC redirects the active power to the DC grid and cannot be considered the slack node for the active power.
Consequently, the slack power constraints \eqref{eq:opf_sq} and \eqref{eq:opf_sp} must be modified:
\vspace{-1pt}
\begin{align}
&\hspace{-5pt} Q_{l}^t = -\sum_{i \in \mathcal{N} }  Q_{i}^t -\sum_{l \in \mathcal{L} \cap \mathcal{L}_{VV} } Q_{l}^t+ Q_{\text{losses}}^{t}, \hspace{10pt} \label{eq:opf_sq2}\\
&\hspace{-5pt} P_{l}^t = -\sum_{i \in \mathcal{N}} P_{i}^t -\sum_{l \in \mathcal{L}} P_{l}^t -\sum_{k \in \mathcal{L}} P_{k}^t - \hspace{-7pt} \sum_{j \in \mathcal{M} \cap \mathcal{M}_{V} } \hspace{-7pt} P_{j}^t + P_{\text{losses}}^{t}  \label{eq:opf_sp2} \vspace{-5pt}
\end{align} 
The objective function of the OPF during \textit{Island} operation is the same as that during \textit{Grid-connected} operation \eqref{eq:opf_obj}. It aims to minimise losses while maintaining the DC slack's SoC at $50\%$. This ensures that the ESS has maximum flexibility to provide voltage control. 

The state machine can only transition to \textit{Island} mode when the grid's state during the \textit{Prepare for island} operation falls within a specific tolerance of the precomputed state in the \textit{Islanded} operation. This will ensure a smooth state transition. 
During the state transition, the state machine issues a command to the synchro-check relay to open the main breaker at the GCP. At the same time, the grid-following IC receives a signal to switch to grid-forming mode, and the ESS converter transitions from power control to voltage control, thereby becoming the new active power slack bus. Due to the delay in the mechanical actions, the main breaker’s opening is intentionally delayed. This ensures that voltage control is active before the grid is physically islanded to prevent any protection interventions.

After the islanding transition, the ramping constraint ensures that the power through the grid-forming converter changes at a sufficiently slow rate. In the \textit{Grid-connected} mode, this power is defined using the active power balance in the DC grid \eqref{eq:opf_form}. However, in the \textit{Island} mode, this expression cannot be used anymore due to the voltage-controllable bus that hosts the ESS\footnote{
The active power balance can only be applied when all but one bus in the DC network regulates the power. The power injected by the voltage-controllable bus can be determined by summing the power injections and line losses. In islanded operation, there are two buses that do not regulate the power: the bus hosting the ESS and the DC bus of the grid-forming IC. Therefore, expression \eqref{eq:opf_form} cannot be used anymore.}.
Therefore, the power injection of the grid-forming IC is defined using the active power balance in the AC grid.

\noindent The constraints \eqref{eq:opf_ramp} and \eqref{eq:opf_form} are replaced by \eqref{eq:opf_ramp2} and \eqref{eq:opf_form2}.
\vspace{-2pt}
\begin{align}
& P_l^{t-1} - \Delta \widehat{P}< P_{l}^t < P_l^{t-1} + \Delta \widehat{P}, \hspace{19pt} \forall \ l \in \mathcal{L}_{VV} , \label{eq:opf_ramp2}\\
& P_{l}^t = -\sum_{i \in \mathcal{N}} P_{i}^t - \hspace{-15pt} \sum_{l \in \mathcal{L} \cap \mathcal{L}_{VV}} \hspace{-10pt} P_{l}^t + P_{\text{losses ac}}^{t}, \hspace{14pt} \forall \ l \in \mathcal{L}_{VV} \label{eq:opf_form2}
\vspace{-2pt}
\end{align}
\subsubsection{State 4: Resynchronisation to the upstream grid}
Once the upper layer grid is restored, the hybrid AC/DC grid can be reconnected. An automatic or manual command initiates the transition to the \textit{Resynchronisation} state. This state has two main objectives.
\begin{enumerate}
    \item To set the power injected by the grid-forming converter to zero to ensure a smooth transition from grid-forming to grid-following mode. This is achieved by updating the objective function as shown in \eqref{eq:opf_obj3}.
\vspace{-2pt}
\begin{align} \vspace{-10pt}
    \eqref{eq:opf_obj} + w_7\lvert \lvert  P_{l,\text{forming}}^t \rvert \rvert + w_8\lvert \lvert  Q_{l,\text{forming}}^t \rvert \rvert \label{eq:opf_obj3}
    \vspace{-2pt}
\end{align}
\item To ensure that the properties of the voltage phasor (magnitude, angle, and frequency) upstream and downstream of the GCP are aligned\footnote{ Only the direct component of the voltage phasors is considered, as both grids may have unbalanced loading conditions.}. The frequency of the grid-forming converter is regulated by a PI controller, which takes the angle difference and regulates the frequency of the grid-forming converter accordingly. 
\end{enumerate}

\vspace{-5pt}

\section{Experimental demonstration}
\label{Sec:experiment}

\subsection{The 27-node hybrid AC/DC microgrid}

\begin{figure*}[t]
    \centering
    \includegraphics[width=\textwidth]{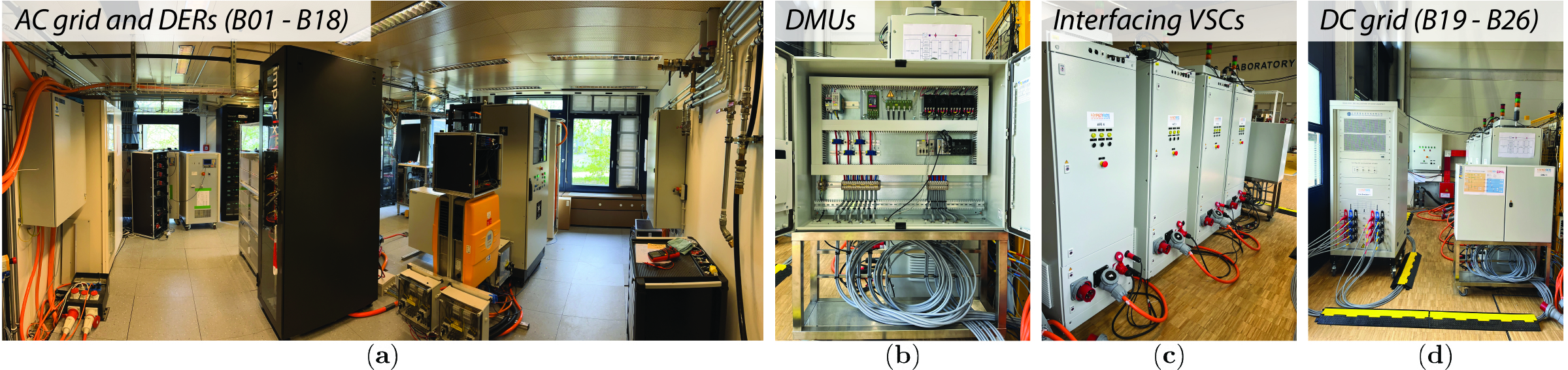}
    \caption{Experimental setup: \textbf{(a)} AC grid and DERs ($B01 - B21$), \textbf{(b)} DC measurement units (\textit{DMU} $1 - 8$), \textbf{(c)} Interfacing converters (\textit{IC} $1-4$), and \textbf{(d)} DC grid ($B19 - B26$).}
    \label{gridphoto} \vspace{-5pt}
\end{figure*}

\begin{figure*}[t]
\begin{minipage}[c]{0.75\textwidth}
\includegraphics[width=\textwidth]{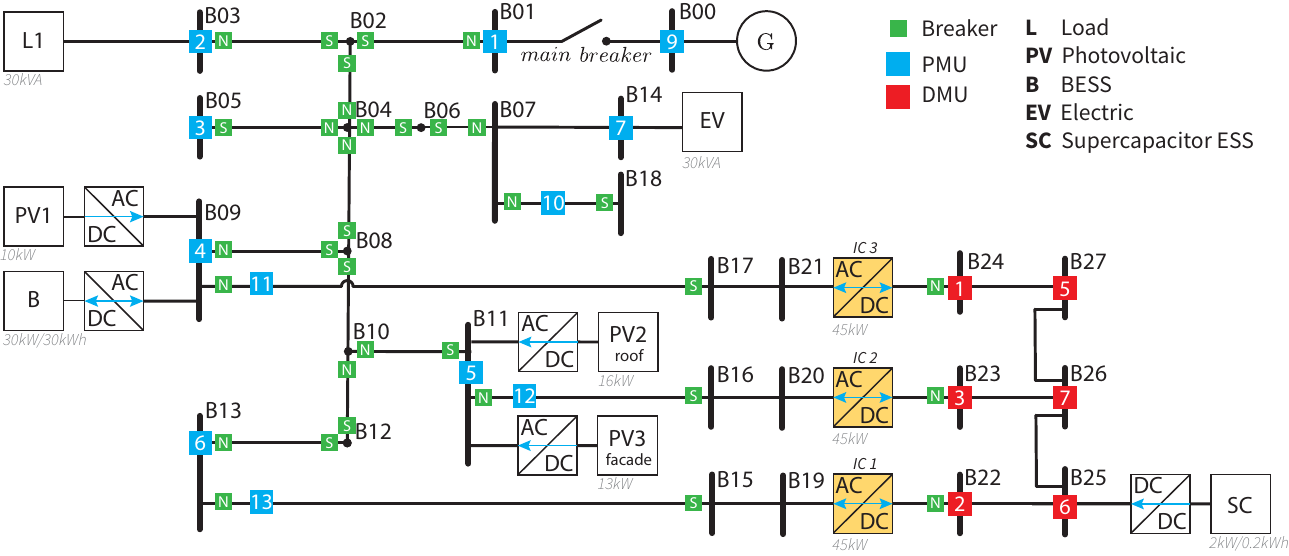}
\captionof{figure}{The hybrid AC/DC low-voltage grid. The resources and their power ratings are indicated. The table provides a summary of the node types. }
\label{fig:Mgrid}
\end{minipage}
\begin{minipage}[c]{0.2\textwidth}
\footnotesize
    \vspace{-0.3cm}
    \begin{tabular}[b]{p{0cm}p{3.1cm}l} 
       & \multicolumn{2}{c}{\textbf{Grid-connected mode}}  \\  \cline{2-3} \cline{2-3}
       & \textbf{Bus Type}     & \textbf{Bus \#}    \\  \cline{2-3} \cline{2-3}
       & Slack-(P,Q)                 & 1                  \\  \cline{2-3} 
       & \textit{PQ}           & 2-18               \\  \cline{2-3} \cline{2-3}
       & $IC_{ac}$ \textit{(Power)}     & 20-21              \\  \cline{2-3} \cline{2-3}
       & $IC_{ac}$ \textit{(Voltage)}   & 19                 \\  \cline{2-3} \cline{2-3}
       & $IC_{dc}$ \textit{(Power)}     & 23-24              \\  \cline{2-3} \cline{2-3}
       & $IC_{dc}$ \textit{(Voltage)}   & 22                 \\  \cline{2-3} \cline{2-3}
       & $P_{dc}$              & 25-27              \\  \cline{2-3} \cline{2-3}
       \vspace{-0.6cm}
       &                             &                   \\
       & \multicolumn{2}{c}{\textbf{Island mode}}   \\  \cline{2-3} \cline{2-3}
       & \textbf{Bus Type}     & \textbf{Bus \#}    \\  \cline{2-3} \cline{2-3}
       & \textit{PQ}           & 1-18               \\  \cline{2-3} \cline{2-3}
       & $IC_{ac}$ \textit{(Power)}     & 20-21              \\  \cline{2-3} \cline{2-3}
       & $IC_{ac}$ \textit{(Forming)} (Slack-Q)  & 19                 \\  \cline{2-3} \cline{2-3}
       & $IC_{dc}$ \textit{(Power)}     & 23-24              \\  \cline{2-3} \cline{2-3}
       & $IC_{dc}$ \textit{(Forming)}  & 22                 \\  \cline{2-3} \cline{2-3}
       & $P_{dc}$              & 26-27              \\  \cline{2-3} \cline{2-3}
       & $V_{dc}$  (Slack-P)     & 25                 \\  \cline{2-3} \cline{2-3}
       \vspace{-0.2cm}
    \end{tabular}
  \label{gridtable} 
\end{minipage}
\vspace{-20pt}
\normalsize
\end{figure*}

The proposed optimal control framework is experimentally demonstrated on the 27-node hybrid AC/DC low-voltage grid developed at the EPFL, Switzerland. The AC part of this experimental hybrid microgrid replicates the CIGRE low-voltage benchmark grid described in \cite{cigre}. It consists of 21 nodes and has a nominal voltage of \SI{400}{V}. The DC part consists of 6 nodes and has a nominal voltage of \SI{720}{V}. Figures \ref{gridphoto} and \ref{fig:Mgrid} present the hybrid grid and its topology. 

Three \SI{45}{\kilo \watt} ICs interconnect the AC and DC grids. Two ICs (IC 2 and IC 3) operate in power mode, while one (IC 1) operates in DC voltage control mode when the hybrid grid is tied to the upstream grid. During island operation, IC 1 transitions to grid-forming mode.
As highlighted in the introduction, this configuration fundamentally differs from islanding demonstrators presented in the literature, where the grid-forming converter is typically directly connected to an ESS (in our case, the DC side of this IC is a network). 
The parameters of the proposed IC loss model presented in \eqref{eq:losses} are: $ [V_0,R_{eq},u,v,w,E_{nom}] = [0.0026, 0.0028, 0.0059, 0.0903, 1.1\text{e}{-4},0.833] \ p.u. $\footnote{The base power is \SI{100}{\kilo \watt}, the AC base voltage is \SI{400}{\volt} and the DC base voltage is \SI{800}{\volt}}. These parameters are extracted from the datasheet of the IGBT module (Semikron SKM 300GB128D) through a least-squared parameter identification process.


\renewcommand{\arraystretch}{1.2}

The hybrid grid hosts several DERs: a \SI{30}{kVA} load emulator representing a household, a \SI{20}{kW}/\SI{20}{kWh} BESS, an electric vehicle charging station, three photovoltaic (PV) plants with a total rated power of \SI{29}{kW} and a supercapacitor ESS rated at \SI{2}{kW}/\SI{0.15}{kWh}. During islanded operation, the supercapacitor transitions from power mode to voltage mode and becomes the slack node for the active power of the entire hybrid grid. The types of nodes in the considered AC/DC microgrid, during both grid-connected and islanded operation, are summarised in Figure \ref{fig:Mgrid}. 

Intentionally, the line \textit{B10-B11} (see Figure \ref{fig:gengrid}) has an ampacity limit of \SI{17}{\ampere}. The power production of the PV connected to bus \textit{B11} will exceed this limit on a sunny day during normal operation when no control is active. To avoid violating the ampacity limit, there are two options: (1) curtailing the PV production, which is not desirable, or (2) using the DC grid to redirect a part of the PV production through uncongested lines.
This realistic operating condition showcases the activation of grid constraints in the OPF problem.
The constraints are active during both islanding and grid-connected operations, and the state machine architecture ensures they are not violated, even during state transitions.

\subsubsection{Sensing and situational awareness}
The formulation of the SC-based OPF requires knowledge of the state of the hybrid grid. This is computed in real-time by a recursive state estimator (SE) originally presented in \cite{willem_SE_exp}. 
The SE is based on an exact linear measurement model and calculates the new state estimates every \SI{100}{\milli \second}. Phasor Measurement Units (PMUs) and DC Measurement Units (DMUs) provided time-synchronised voltage and current measurements. 
The PMUs use the enhanced Interpolated Discrete Fourier Transform (e-IpDFT) algorithm \cite{romano2016dft} to extract the nodal voltage and current phasors. These measurement units are P-class devices that conform to the IEEE standard C37.118.
The DMUs extract the DC nodal voltage and current using an averaging function. 
The measurements are time-aligned by the Phasor Data Concentrator (PDC) \cite{asjaPDC} before being sent to the SE.




\subsubsection{Synchro-check relay}
The synchro-check verifies that the voltage phasors across the main breaker (between buses \textit{B00} and \textit{B01}, as shown in Figure  \ref{fig:Mgrid}) are synchronised prior to closing. It is specifically designed for this experiment and implemented on an NI CompactRio system. The voltage phasors are extracted using the same e-IpDFT of the PMUs \cite{romano2016dft}. Only the direct sequence components of the voltage phasors are considered. When the difference between the direct sequence voltages across the main breaker (phasor angles, magnitudes, and frequencies) is within the predefined tolerances, the synchro-check will send a command to the state machine to allow resynchronising the two grids. The tolerances used in the experimental demonstration are summarized in Table 
\ref{tab:synchrocheck}. 
\begin{table}[h] 
    \centering
    \begin{tabular}{c c}
        \textbf{Phasor quantity} & \textbf{Tolerance} \\
        \hline \hline
        Magnitude &  \SI{5}{\volt}\\
        \hline
        Angle &  \SI{2}{\degree}\\
        \hline
        Frequency &  \SI{0.02}{\hertz}\\
        \hline
    \end{tabular}
    \caption{Tolerances of the synchro-check relay}
    \label{tab:synchrocheck} 
\end{table}
\subsubsection{Centralized control architecture}
\begin{figure}[h] \vspace{-10pt}
\includegraphics[width=\linewidth]{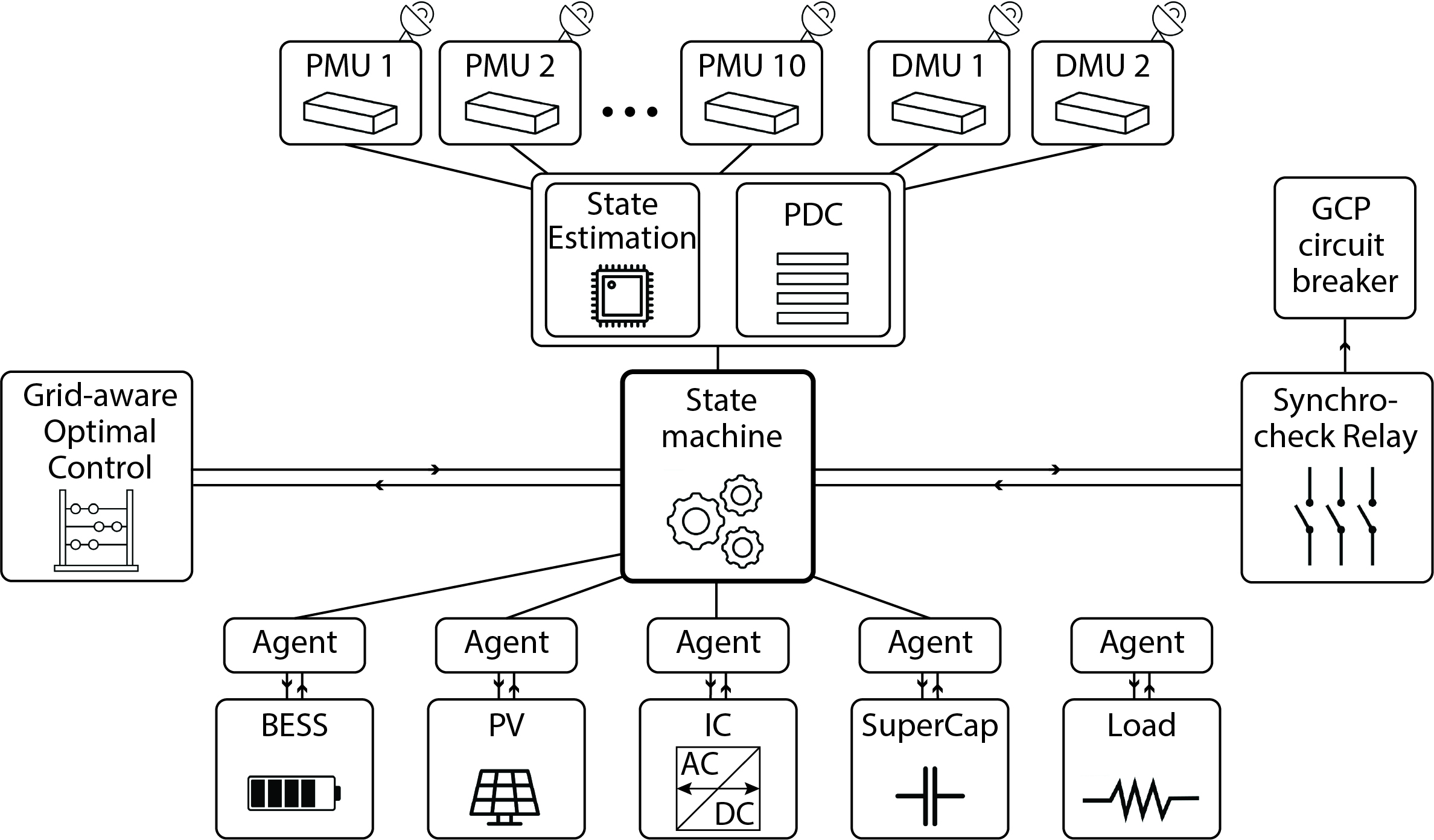}
\captionof{figure}{Architecture of the islanding and resynchronization control framework.}
\label{fig:Control_architecture} \vspace{-20pt}
\end{figure}

The centralized control architecture of the grid-aware islanding and resynchronisation framework is illustrated in Figure \ref{fig:Control_architecture}. The state machine is the backbone of the control architecture and facilitates all communication with the state estimator, the optimal control algorithm, the synchro-check relay, and the DERs. Every DER has a dedicated resource agent that translates the resource-specific communication protocol (Modbus, CAN, etc.) to a standardized UDP frame and ensures the correct start-up, shutdown, and error-handling procedure. The load emulator does not receive references from the state machine, as it is uncontrollable and autonomously tracks the consumption profile of a typical cluster of households.

\vspace{-5pt}
\subsection{Case study}
\label{sec:usecase}
\begin{figure}[!h] \vspace{-6pt}
  \includegraphics[width=\linewidth]{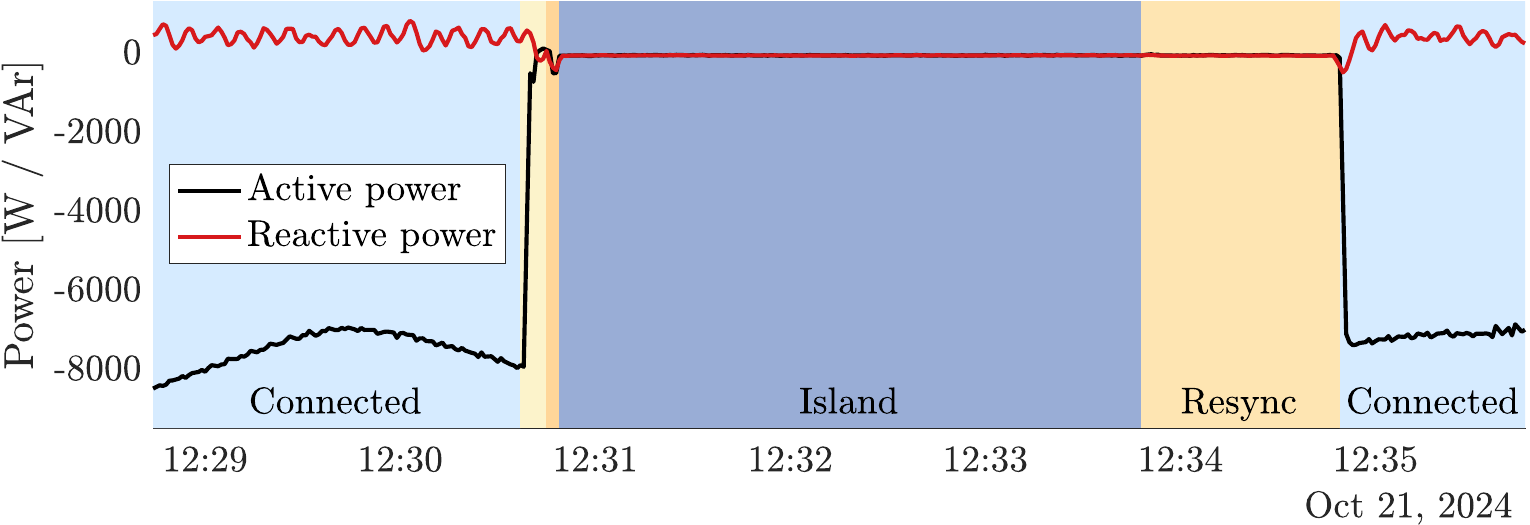}
  \caption{Active and reactive power of the GCP (Bus \textit{B01}).}
  \label{fig:GCP_power} \vspace{-8pt}
\end{figure}

\begin{figure}[!h] \vspace{-8pt}
  \includegraphics[width=\linewidth]{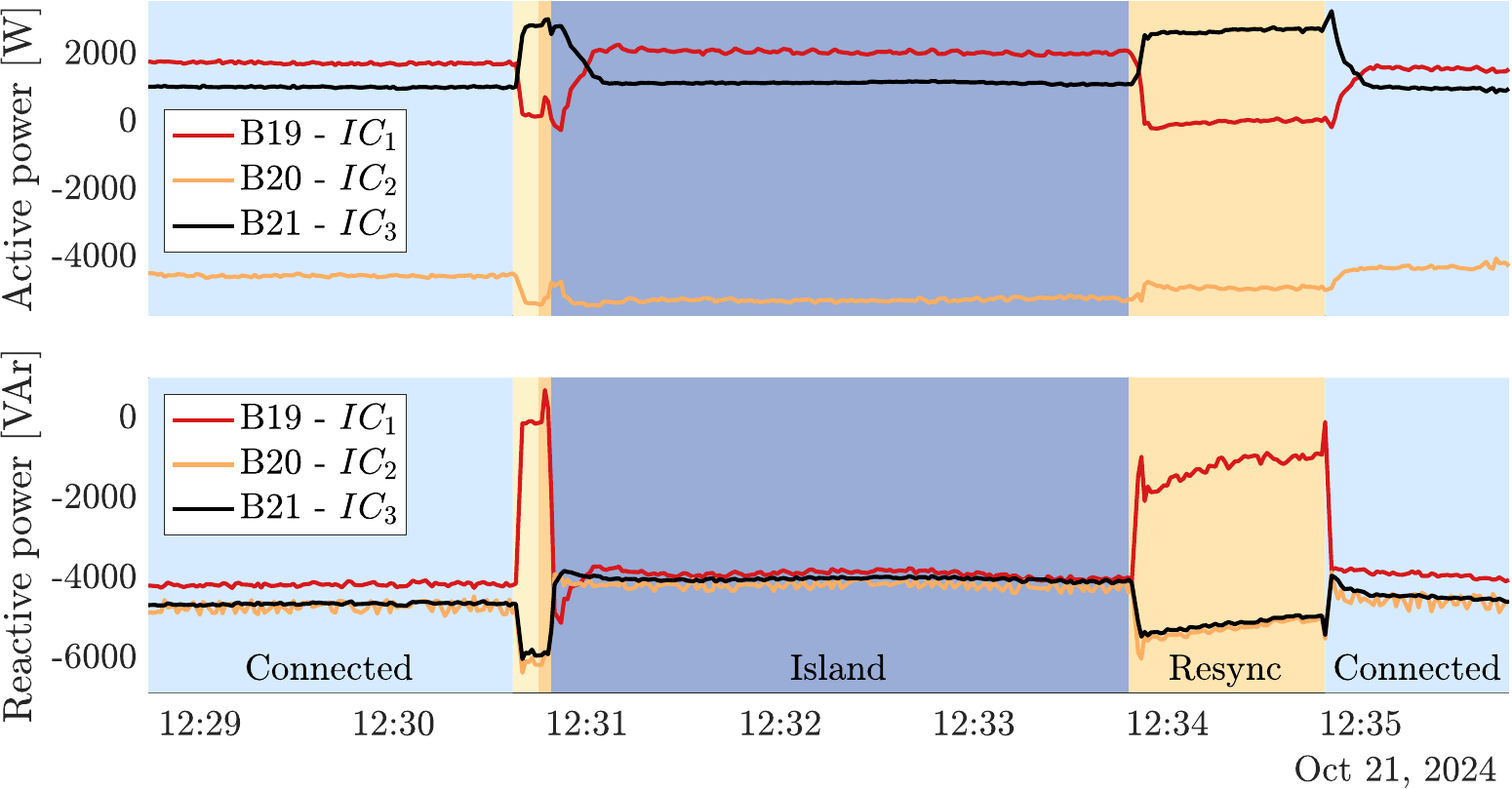}
  \caption{Active and reactive power of IC 1-3.}
  \label{fig:IC_power}  \vspace{-8pt}
\end{figure}

\begin{figure}[!h] \vspace{-8pt}
  \includegraphics[width=\linewidth]{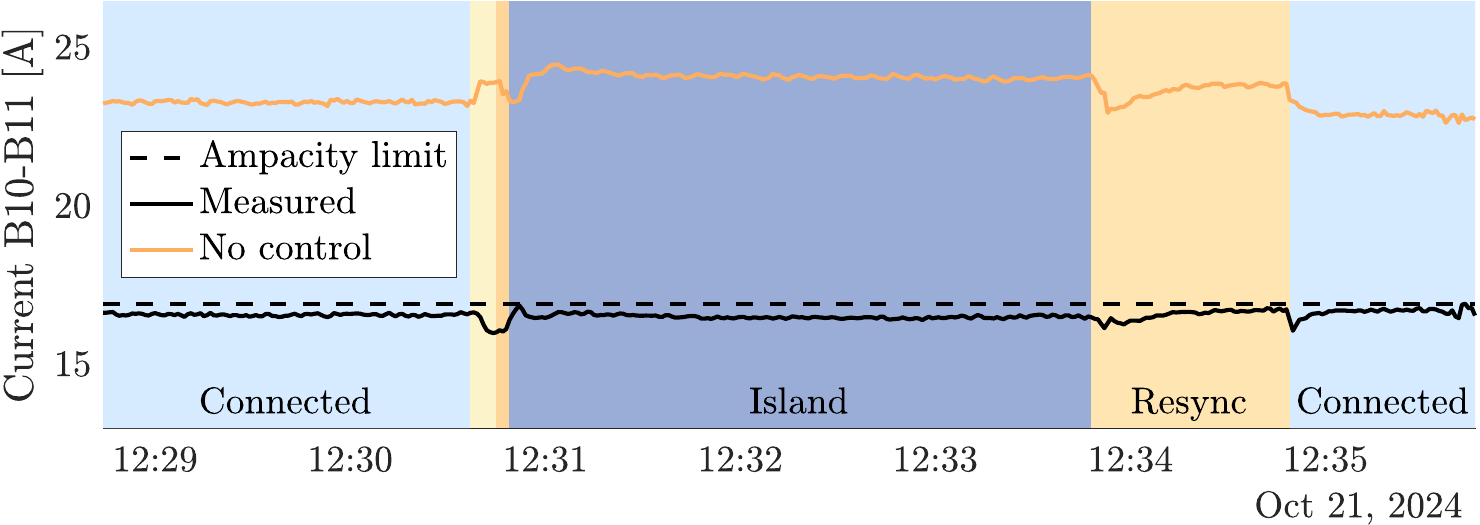}
  \caption{Current flow in the congested line \textit{B10-B11}. The ampacity limit is marked by the dashed line, and the current flow, under 'no control', is calculated a posteriori and presented in orange.}
  \label{fig:Current}  \vspace{-6pt}
\end{figure}

\begin{figure}[!h] \vspace{-6pt}
  \includegraphics[width=\linewidth]{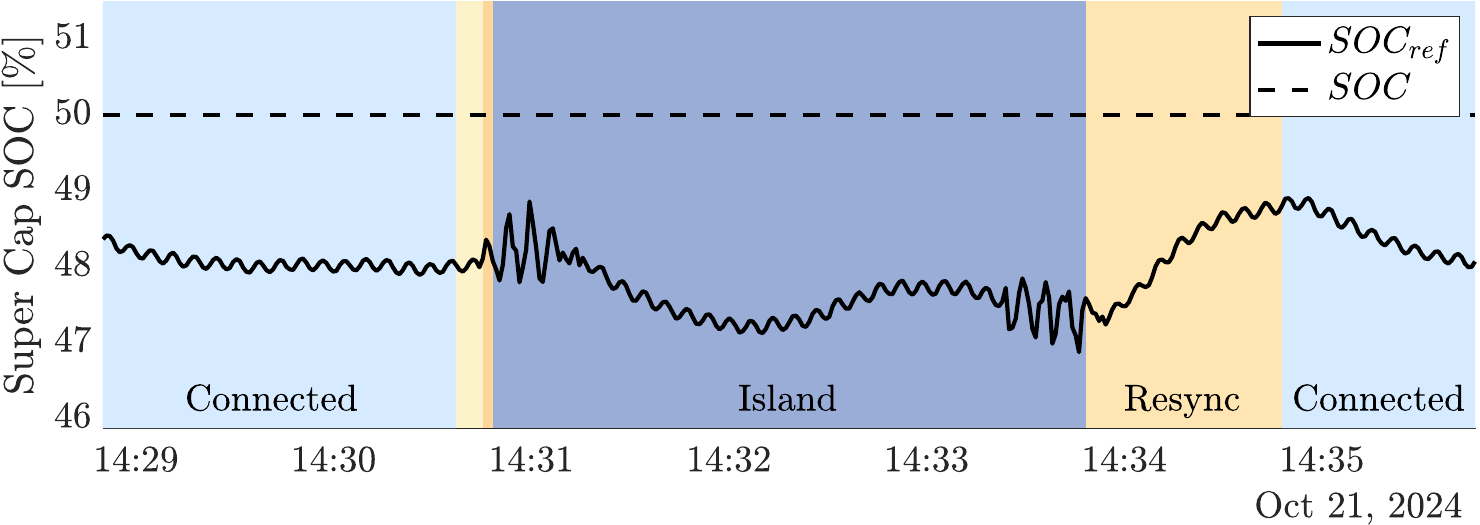}
  \caption{SoC of the supercapacitor ESS. This resource transitions from power mode to voltage mode during the island operation.}
  \label{fig:soc} \vspace{-10pt}
\end{figure}

\vspace{-5pt}
The experiment, conducted on October 21, 2024, demonstrates and validates the islanding and resynchronization framework. The results are shown in Figures \ref{fig:GCP_power}—\ref{fig:soc}. The operation modes are distinctly indicated on all figures by the shaded areas: light blue represents \textit{Grid-connected} mode, yellow represents \textit{Prepare for island} mode, dark blue represents \textit{Island} mode, and dark yellow represents the 	\textit{Resynchronisation} mode.
At the start of the experiment, the grid operates in \textit{Grid-connected} mode. Approximately \SI{8}{\kilo \watt} is transferred through the GCP, see Figure \ref{fig:GCP_power}. The ICs redirect power through the DC grid (see Figure \ref{fig:IC_power}) to avoid exceeding the ampacity limit of line \textit{B10-B11} (see Figure \ref{fig:Current}). The current rating in the case of 'no-control' is calculated a posteriori and clearly shows a grid constraint violation.

Following a manual action (at $t = 13{:}30{:}37$), the state machine transitions to \textit{Prepare for island} mode. The active and reactive power flows through the GCP and the IC 1 are reduced to zero (see Figure \ref{fig:GCP_power} and \ref{fig:IC_power}) to ensure a minimal power transient when the breaker opens. This is achieved by the updated objective function given in \eqref{eq:opf_obj2}.

The state machine can only transition to \textit{Island} mode when the transition criteria are satisfied to ensure grid state consistency before and after the manoeuvre (see Section \ref{sec:architecture}). This is archived by solving two OPF problems: one for the island topology and another for the grid-connected topology, where the grid constraints of one are the objective function of the other \eqref{eq:opf_obj2}. At the state transition, the supercapacitor at bus \textit{B25} first switches to voltage mode to regulate the DC voltage. Next, IC 1 shifts to grid-forming mode, providing voltage control to the AC grid. Finally, the main circuit breaker opens after a fixed \SI{2}{\second} delay.
Once the grid is islanded, the ramping constraints in the grid-forming IC are active and limit the rate of change of its active power injection. The power transferred through the ICs reaches values similar to before the island operation. 
Furthermore, Figure \ref{fig:Current} shows that throughout the transition,  the current flow in line \textit{B10-B11} never violates its ampacity limit.

\begin{figure}[!h] \vspace{-6pt}
  \includegraphics[width=\linewidth]{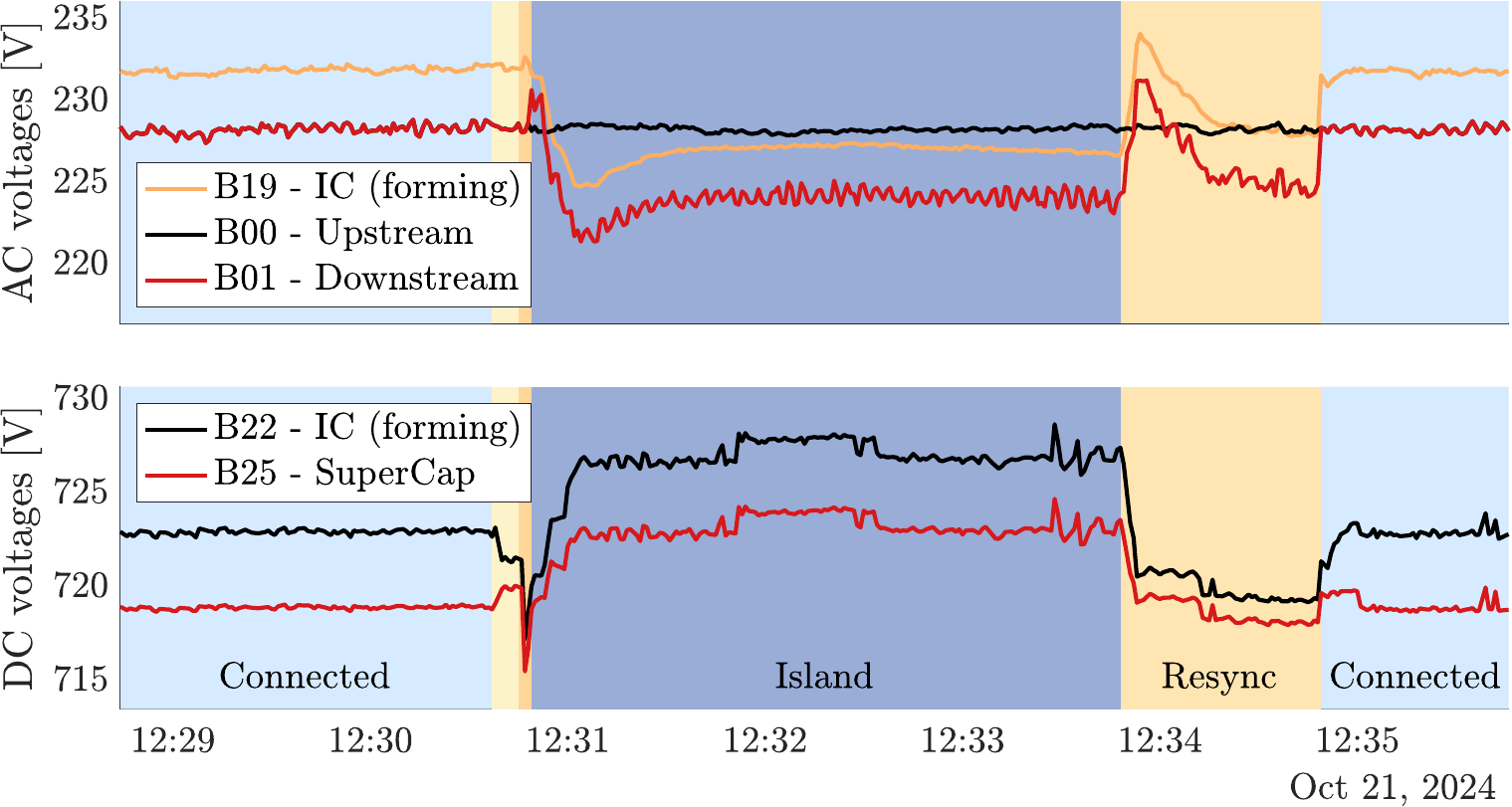}
  \caption{Nodal voltage profile of certain nodes in the AC and DC network.}
  \label{fig:voltage} \vspace{-8pt}
\end{figure}

\begin{figure}[!h] \vspace{-8pt}
  \includegraphics[width=\linewidth]{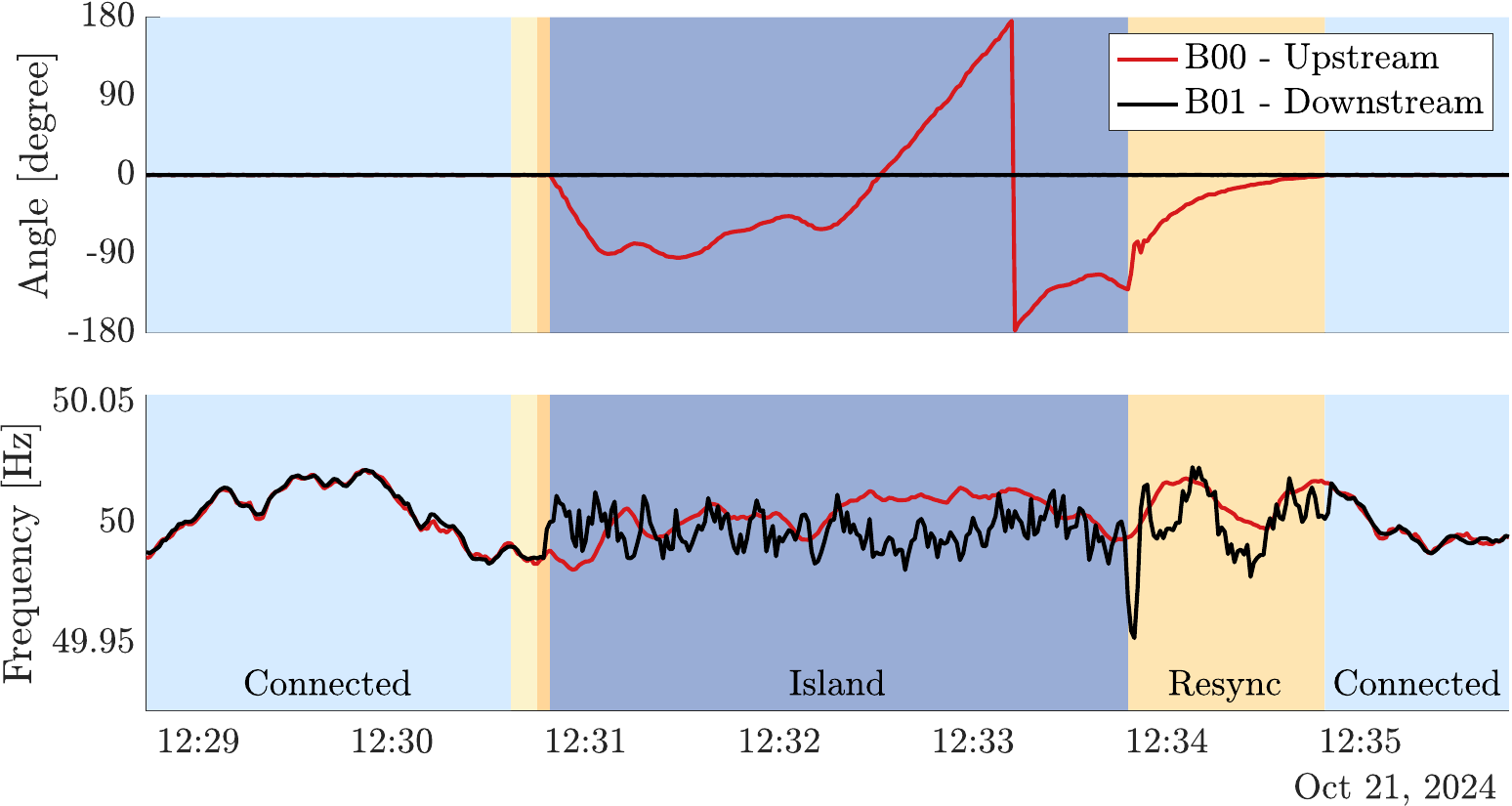}
  \caption{Angle and frequency of the upstream grid and downstream grid. The downstream angle is chosen as the reference.}
  \label{fig:angle_freq} \vspace{-7pt}
\end{figure}

The grid operates in island mode for approximately \SI{3}{\minute}. As illustrated in Figure \ref{fig:voltage}, both the AC and DC voltages remain stable around their nominal values. Figure \ref{fig:angle_freq} indicates that the frequency fluctuates around \SI{50}{\hertz} with maximum deviations of $\pm$ \SI{0.03}{\hertz}.
When the synchro-check relay detects that the upper layer grid is restored (initiated here by a manual action), the state transitions to \textit{Resynchronisation} mode. As shown in Figure \ref{fig:IC_power}, the power through the grid-forming IC 1 reduces to zero. This power cannot be directly regulated, as grid-forming operation only controls voltage magnitude and frequency\footnote{ The voltage phasor angle is considered as a bias to the frequency in the converter control.}. Therefore, the OPF regulates the setpoints of the other controllable DERs in order to bring this power to zero. 

Figure \ref{fig:angle_freq} shows the angle and the frequency over the full islanding and resynchronisation manoeuvre. During the island operation, the angle of the upstream network drifts away from the downstream grid angle (the reference angle). At the start of the \textit{Resynchronization} mode, an angle difference of \SI{150}{\degree} is observed. The state machine can only transition back to \textit{Grid-connected} mode when the phasor differences are within the predefined tolerance given in Table \ref{tab:synchrocheck}.
Therefore, a PI controller tracks this angle by controlling the frequency of the grid-forming converter. It takes approximitelly \SI{50}{\second} to achieve the required tolerance of \SI{2}{\degree} (see Figure \ref{fig:angle_freq}). The PI controller on the angle updates its control action every \SI{100}{\milli \second}. 
Once the tolerances are met, the grid can synchronise. First, the main breaker closes, then the grid-forming IC transitions to grid-following operation, and finally, the supercapacitor transitions to power mode.

The voltage profiles of the AC and DC networks are shown in Figure \ref{fig:voltage}. During the experiment, the AC voltage exhibits a maximum deviation of approximately  \SI{7}{\volt}, which corresponds to a $3\%$ deviation from the nominal value. The DC voltage experiences a deviation of around \SI{8}{\volt}, corresponding to $1\%$ of the nominal DC voltage.
This demonstrates that the islanding and resynchronisation framework, which incorporates the state machine and the grid-aware OPF, effectively mitigates the strong transient behaviour that is typically observed.
The SoC of the Supercap ESS is shown in Figure \ref{fig:soc}. The reference SoC is $50\%$, indicated by the dashed line. The control mode of the supercapacitor converter changes from power mode to voltage mode during the islanding manoeuvre. The SoC remains close to its reference value, maximizing the flexibility of voltage control during islanded operation.

\begin{figure}[!h]  \vspace{-5pt}
  \includegraphics[width=\linewidth]{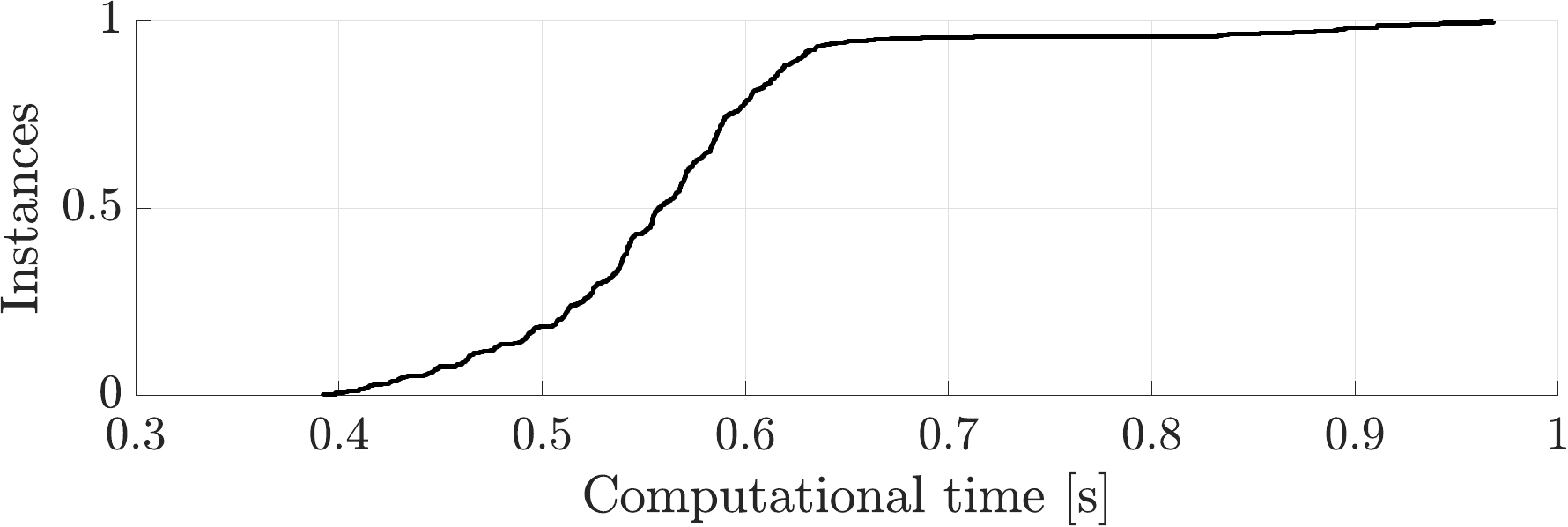}
  \caption{Cumulative density function of the computational time of the OPF program.}
  \label{fig:time}  \vspace{-7pt}
\end{figure}

Finally, the computational time of the grid OPF computation is presented in Figure \ref{fig:time} as a cumulative density function. It is consistently below \SI{1}{\second}, allowing the optimal control framework to compute the optimal set points of the resources at one-second intervals.

\vspace{-5pt}

 \vspace{-8pt}
\section{Conclusions}

The paper presents an optimal control framework for the islanding and resynchronisation of hybrid AC/DC networks. The framework is based on a unified load flow model that integrates both the AC and DC networks, as well as the ICs ($\geq1$). 
While earlier work from the authors presented a unified load flow model for grid-following ICs, this work extends the model to accommodate grid-forming operations. These grid-forming ICs fundamentally change the network's topology, as the power needed to maintain the active power balance is injected by this grid-forming IC but supplied by the DC grid. 
As a result, two distributed slack nodes are introduced: one for reactive power in the AC grid (at the AC side of the grid-forming IC) and another for active power in the DC grid.
This configuration enhances the flexibility and resilience of the system.
The unified load flow model is linearized using sensitivity coefficients and incorporated into the OPF framework to model grid constraints.

The optimal control framework is experimentally validated on an actual 27-node hybrid AC/DC network.
Certain grid constraints are binding throughout the islanding and resynchronisation manoeuvre but are never violated during the operation or state transitions, demonstrating the effectiveness of the grid-aware OPF framework.
The voltage deviations remain below $3\%$ of the nominal value during state transitions. Additionally, the grid-forming IC adjusts its frequency during resynchronization and tracks the angle difference across the main breaker to less than \SI{2}{\degree} before synchronizing with the upstream grid.
Finally, the framework's computational efficiency enables a one-second refresh rate, allowing it to effectively handle rapid changes in grid state caused by state transitions and renewable energy intermittency.

\vspace{-5pt}


\bibliographystyle{IEEEtran} \vspace{-5pt}
\bibliography{IEEEexample.bib} \vspace{-5pt}

\end{document}